\documentclass[reprint,aps,prd,onecolumn,notitlepage,nofootinbib,preprintnumbers,superscriptaddress]{revtex4-1}
\usepackage{amsmath}
\usepackage{amsfonts}
\usepackage{amssymb}
\usepackage{latexsym}
\usepackage{enumerate}
\usepackage{color,xcolor}
\usepackage{graphicx}
\usepackage{bm}
\usepackage{slashed}
\usepackage{epsfig}
\usepackage[english]{babel}
\usepackage{times}
\usepackage{comment}
\usepackage{hyperref}
\usepackage{epstopdf}
\usepackage{extarrows}
\usepackage{lipsum}
\usepackage{tikz}
\tikzset{every picture/.style={line width=0.75pt}} 

\def\aF{\tilde{F}}
\def\d{\mathrm{d}}
\def\e{\mathrm{e}}
\def\I{\mathrm{i}}

\def\vB{\mathbf{B}}
\def\hvB{\hat{\mathbf{B}}}
\def\tvB{\tilde{\mathbf{B}}}
\def\vE{\mathbf{E}}

\def\vS{\mathbf{S}}
\def\vb{\mathbf{b}}
\def\ve{\mathbf{e}}
\def\vk{\mathbf{k}}
\def\hvk{\hat{\mathbf{k}}}
\def\vx{\mathbf{x}}

\def\vv{\mathbf{v}}
\def\vp{\mathbf{p}}
\def\vn{\mathbf{n}}
\def\hvn{\hat{\mathbf{n}}}
\def\vm{\mathbf{m}}
\def\hvm{\hat{\mathbf{m}}}
\def\hvl{\hat{\mathbf{l}}}
\def\hvi{\hat{\mathbf{i}}}

\begin{document}
\title{Compton scattering in Bandos-Lechner-Sorokin-Townsend nonlinear electrodynamics}
\author{Yang Shi}
\affiliation{School of Physics and Electronic Science, East China Normal University, Shanghai 200241, China\\ \vspace{0.2cm}}
\author{Towe Wang}
\email[Electronic address: ]{twang@phy.ecnu.edu.cn}
\affiliation{School of Physics and Electronic Science, East China Normal University, Shanghai 200241, China\\ \vspace{0.2cm}}
\date{\today\\ \vspace{1cm}}
\begin{abstract}
The nonlinear electrodynamics proposed by Bandos, Lechner, Sorokin and Townsend is a remarkable theory that unifies Maxwell, Bialynicki-Birula and ModMax theories, which are known theories invariant under conformal transformations and electromagnetic duality transformations. In the Bandos-Lechner-Sorokin-Townsend nonlinear electrodynamics, we calculate the energy flux density, dispersion relations, refractive indices, phase and group velocities of plane waves as well as the changes of the photon wavelength in the Compton scattering process in the presence of a constant uniform electromagnetic background. Our results are useful for testing and constraining this new theory of nonlinear electrodynamics.
\end{abstract}

\maketitle
\section{Introduction}\label{sec:intro}
Maxwell electrodynamics is a successful theory describing electromagnetic phenomena with high precision. It is broadly accepted in both physical theories and engineering practices. Nevertheless, together with theoretical calculations of quantum electrodynamics, ultra-precise experiments are performed or undergoing to test Maxwell electrodynamics in extreme circumstances. Some of them seek for evidences of the vacuum birefringence \cite{Della_Valle_2014,EJLLI20201,Ataman:2023gin,Ataman:2025,Shen_2018}. If such evidences are confirmed, there will be two possible ways to explain them: either Maxwell electrodynamics needs classical (non-quantum) corrections, or Maxwell electrodynamics is an effective theory to be extended to incorporate nonlinear quantum processes. For both possibilities, more investigations on nonlinear electrodynamics are necessary.

In history, studies of the vacuum polarization effect led to Heisenberg-Euler electrodynamics \cite{heisenberg2006consequences}, which can be obtained also from the one-loop correction of quantum electrodynamics. Another theory of nonlinear electrodynamics is Born-Infeld electrodynamics \cite{doi:10.1098/rspa.1934.0059}, which was invented to eliminate the divergence of self-energy of a free electron and then revived in the superstring theory \cite{FRADKIN1985123,Gibbons_2001}.

As is well known, Maxwell electrodynamics is invariant under both conformal transformations and $SO(2)$ electromagnetic duality transformations \cite{jackson_classical_1999}. It is less known that Bialynicki-Birula electrodynamics \cite{Bialynicki-Birula1983}, which was first found as a strong-field limit of Born-Infeld electrodynamics, is invariant under both conformal transformations and $SL(2,R)$ electromagnetic duality transformations\cite{Bialynicki-Birula:1992rcm}. Recently, a new theory of electrodynamics, modified Maxwell (ModMax) electrodynamics, was discovered \cite{Bandos_2020}. It preserves both of the aforementioned symmetries of Maxwell electrodynamics. As a bridge to ModMax electrodynamics, a generalized Born-Infeld electrodynamics was also proposed in Ref. \cite{Bandos_2020}. This theory is duality-invariant but not conformally invariant, just like Born-Infeld electrodynamics. Following Ref. \cite{shi2023transverse}, we will refer to it as Bandos-Lechner-Sorokin-Townsend (BLST or BLeST) electrodynamics. Remarkably, BLST electrodynamics unifies Maxwell, Born-Infeld, Bialynicki-Birula and ModMax theories \cite{Bandos_2020,Bandos_2021,Sorokin_2022,shi2023transverse}, and it can be constructed as $T\bar{T}$-like deformations of Born-Infeld electrodynamics, ModMax electrodynamics or Maxwell electrodynamics \cite{Conti_2018,Ferko_2022,BABAEIAGHBOLAGH2022137079,ferko2023dualityinvariant}.

In the literature, there are many works on Born-Infeld electrodynamics and ModMax electrodynamics, such as transverse electric waves\cite{Ferraro_2007,MANOJLOVIC2020168303}, birefringences \cite{Russo_2023,Neves_2023,mezincescu2024hamiltonian} and Compton scattering \cite{PhysRevD.104.015006, Lechner_2022}. In contrast, BLST electrodynamics has been less explored. In Ref. \cite{shi2023transverse} we tried to fill this gap by computing the transverse electric waves in BLST electrodynamics, thereby making it possible to test this theory in a parallel plate waveguide. In Ref. \cite{SHI_Blackbody}, we delved into the blackbody radiation spectrum and thermodynamic quantities in BLST electrodynamics. In the present article, we will continue to design a configuration to test BLST electrodynamics with the Compton scattering experiment in a constant uniform magnetic background.

The structure of this article is as follows. For theoretical preparation, in Section \ref{sec:framework} we will start with a general Lagrangian of nonlinear electrodynamics exclusively depending on two real Lorentz-invariant scalars, $S = -F_{\mu\nu}F^{\mu\nu}/4$ and $P = -F_{\mu\nu}\aF^{\mu\nu}/4$, where the powers of $P$ are even to avoid violation of the charge-parity (CP) symmetry. Perturbatively expanding the Lagrangian around an arbitrary constant uniform electromagnetic background, we will derive the linearized field equations for plane wave perturbations, from which the energy density, the energy flux density and the weak energy condition will also be worked out. In Section \ref{sec:DR}, we will find modified dispersion relations of the plane waves and calculate some measurable physical quantities such as the refractive index, the phase velocity and the group velocity following the method in Ref. \cite{PhysRevD.104.015006}. Rather than being restricted to a purely magnetic background as in Ref. \cite{PhysRevD.104.015006}, we will consider a general background of combined electric and magnetic fields in Subsection \ref{subsec:Comb}. The special situations with a pure electric or magnetic background field will be investigated further in Subsection \ref{subsec:Pure}. Based on the dispersion relations of photons in nonlinear electrodynamics, we will turn to Compton scattering in Section \ref{sec:scatt} and write down a general formula for the wavelength of scattered waves. We will also give a simplified formula for the wavelength shift when phase velocities are close to $c$. In Section \ref{sec:inBLST}, we will apply our results to BLST electrodynamics in a region of a constant magnetic field, and then design a specific Compton scattering experiment to constrain parameters of this theory. After presenting dispersion relations in Subsection \ref{subsec:DRinBLST}, we will demonstrate in Subsection \ref{subsec:fluxinBLST} that usually the wave vector and the wave energy flux (light ray) are along different directions. To ensure that they are in the same direction and accessible to experiment, in Subsection \ref{subsec:scattinBLST}, we will pay attention to the Compton scattering process in a plane perpendicular to the magnetic field. We will discuss the implications of our research and conclude in Section \ref{sec:conclusion}.

Throughout this article, we adopt the metric convention ($+,-,-,-$) and work in natural units with $c=\hbar=1$. In such units, the momentum and energy of a photon are equal to the wave vector $\vk$ and the circular frequency $\omega$, respectively, while the wavelength of the photon is $\lambda=2\pi/|\vk|$. We assume the parameters of BLST electrodynamics satisfy $T>0$ and $\gamma \geqslant 0$. A unit vector is represented by a letter marked with a hat, e.g., $\hvk=\vk/|\vk|$.

In our conventions, the background fields $\vE,\vB$ and the plane wave amplitudes $\ve_0,\vb_0$ are real vectors, but the plane wave fields $\ve=\ve_0\e^{\I(\vk\cdot\vx-\omega t)}$, $\vb=\vb_0\e^{\I(\vk\cdot\vx-\omega t)}$ are complex. If the readers are not familiar with the complex form of plane waves, they can simply replace the unit complex functions with trigonometric functions, for instance, $\ve=\ve_0\sin(\vk\cdot\vx-\omega t)$, $\vb=\vb_0\sin(\vk\cdot\vx-\omega t)$ throughout this article.

\section{Field equations and continuity equation}\label{sec:framework}
In this section, we will introduce some notations and the framework of our work. We begin with the most general form of Lagrangian in $D=4$ Minkowski spacetime, which does not include derivatives of field strength $F_{\mu\nu} = \partial_{\mu}A_{\nu} - \partial_{\nu}A_{\mu}$ to avoid ghosts, as a function of the Lorentz and gauge invariants
\begin{subequations}
\begin{align}
   S &= -\frac{1}{4}F_{\mu\nu}F^{\mu\nu} = \frac{1}{2}\left( \vE^2-\vB^2 \right)    ,\\
   P &= -\frac{1}{4}F_{\mu\nu}\aF^{\mu\nu} = \vE \cdot \vB,
\end{align}
\end{subequations}
where $\aF^{\mu\nu}$ is the Hodge dual of the field strength $F^{\mu\nu}$. Therefore we can write the Lagrangian as $\mathcal{L}(S,P)$ formally. It can go back to Maxwell electrodynamics if we choose $\mathcal{L} = S$. We will restrict ourselves to CP-invariant theories, or equivalently, theories with time-reversal invariance, so we assume the powers of $P$ in Lagrangian $\mathcal{L}(S,P)$ are even throughout this article.

Any nonlinear electrodynamics with Lagrangian of the form $\mathcal{L}(S,P)$ has a trivial solution, a constant uniform electromagnetic field. We are interested in linear perturbations to the background solution, which can also be thought of as plane waves propagating in the background electromagnetic field. Accordingly, we decompose the full electromagnetic potential $A^{\mu}$ into two parts, the background potential $A_0^{\mu}$ and the plane wave $a^{\mu}$ as a perturbation. Then the field strength can be decomposed in the same way, $F^{\mu\nu} = F_0^{\mu\nu} + f^{\mu\nu}$, in which $F_0^{\mu\nu}$ corresponds to the field strength associated with the electric and magnetic background fields, $F_0^{0i}=E_i$, $F_0^{ij}=\epsilon^{ijk}B_k$, whereas $f^{\mu\nu}$ is the electromagnetic field strength tensor of the propagating excitation, $f^{0i}=e_i$, $f^{ij}=\epsilon^{ijk}b_k$. In this way, by expanding the total Lagrangian $\mathcal{L}(S,P)$ around an intense background field, we get the second-order Lagrangian of a relatively weak plane wave \cite{Bialynicki-Birula1983,P_rez_Garc_a_2023}
\begin{equation}\label{eq:Lag}
    \mathcal{L}^{(2)} = -\frac{\mathcal{L}_S}{4} \left(
        f_{\mu\nu}f^{\mu\nu} + \ell_0 f_{\mu\nu}\tilde{f^{\mu\nu}}
        - Q^{\mu\nu\kappa\lambda}\frac{1}{2}f_{\mu\nu}f_{\kappa\lambda}
    \right) ,
\end{equation}
where $Q^{\mu\nu\kappa\lambda}$ is a tensor made of background field strength. This tensor is defined as
\begin{align}
    Q^{\mu\nu\kappa\lambda} &= \ell_1 F_0^{\mu\nu}F_0^{\kappa\lambda} + \ell_2 \tilde{F}_0^{\mu\nu} \tilde{F}_0^{\kappa\lambda}
    + \ell_3 F_0^{\mu\nu}\tilde{F}_0^{\kappa\lambda} + \ell_3 \tilde{F}_0^{\mu\nu}F_0^{\kappa\lambda}.
\end{align}
Here we have used the notations $\ell_0=\mathcal{L}_P/\mathcal{L}_S$, $\ell_1=\mathcal{L}_{SS}/\mathcal{L}_S$, $\ell_2=\mathcal{L}_{PP}/\mathcal{L}_S$, $\ell_3=\mathcal{L}_{SP}/\mathcal{L}_S$, and the subscripts of Lagrangian $\mathcal{L}$ denote partial derivatives of the Lagrangian. These coefficients are evaluated at the background level, while the strength of plane wave is on the perturbative level. We assume that $\mathcal{L}_S \neq 0$, because otherwise the linearized field equation does not propagate two independent polarization wave modes \cite{Russo_2023}, and it is also the result of the weak energy condition to be discussed at the end of this section.

Corresponding to the above Lagrangian \eqref{eq:Lag}, the linearized field equations and the Bianchi identities are of the form
\begin{align}\label{eq:field}
    \partial_{\mu}\left[ f^{\mu\nu} + \ell_0 \tilde{f}^{\mu\nu} - \frac{1}{2}f_{\kappa\lambda}Q^{\mu\nu\kappa\lambda} \right]
    =  0
    ,\quad
    \partial_{\mu}\tilde{f}^{\mu\nu} = 0.
\end{align}
In this article, we will make use of them to study plane waves in an intense constant uniform electromagnetic background. To proceed, let us convert them into the vector form in terms of the background field $(\vE,\vB)$ and the plane wave $(\ve,\vb)$,
\begin{subequations}\label{eq:fieldeq}
\begin{align}
    &\nabla\times\ve+\partial_t\vb = 0   ,\quad
    \nabla\cdot\vb = 0,
    \label{eq:fieldeqA}\\
    &\left(\ell_3 \vB+\ell_1 \vE\right)\cdot\nabla(\vE\cdot\ve-\vB\cdot\vb)
    +\left(\ell_2\vB +\ell_3 \vE\right)\cdot\nabla(\vB\cdot\ve+\vE\cdot\vb)
    +\nabla\cdot \ve
    = 0,
    \label{eq:fieldeqB}\\
    &\left(\ell_3 \vE-\ell_1\vB \right)\times \nabla(\vE\cdot\ve-\vB\cdot\vb)
    +\left(\ell_2 \vE-\ell_3\vB \right)\times \nabla(\vB\cdot\ve+\vE\cdot\vb)
    \nonumber\\
    &+\nabla\times \vb
    = \partial_t \ve
    +\left(\ell_3\vB +\ell_1 \vE\right) \partial_t (\vE\cdot\ve
    -\vB\cdot\vb)
    +\left(\ell_2\vB +\ell_3 \vE\right) \partial_t (\vB\cdot\ve+\vE\cdot\vb),
    \label{eq:fieldeqC}
\end{align}
\end{subequations}
where real constant vectors $\vE=(E_1,E_2,E_3)$, $\vB=(B_1,B_2,B_3)$ are electric and magnetic background fields, and nonconstant vectors $\ve=(e_1,e_2,e_3)$, $\vb=(b_1,b_2,b_3)$ are electric and magnetic components of the plane wave.

Following the procedure in Maxwell electrodynamics, we can take the inner product of $\mathcal{L}_S\ve$ and \eqref{eq:fieldeqC}, and then transform the result into the continuity equation
\begin{equation}\label{eq:cont}
   \partial_t w + \nabla\cdot\vS = 0.
\end{equation}
In this equation, the energy density $w$ and the energy flux density $\vS$ are defined as
\begin{subequations}
    \begin{align}
        w &= \frac{1}{2}\left( K_{ij}e_i e_j + P_{ij}b_i b_j \right),
        \label{eq:eng_den}\\
        \vS &= \mathcal{L}_S \ve\times\left[
        \vb
        +(\ell_1 \vB - \ell_3 \vE)(\vE\cdot\ve - \vB\cdot\vb)
        +(\ell_3 \vB - \ell_2 \vE)(\vB\cdot\ve + \vE\cdot\vb)
        \right],
        \label{eq:eng_flu}
    \end{align}
\end{subequations}
with the coefficient matrices \cite{PhysRevD.104.015006}
\begin{subequations}
    \begin{align}
        K_{ij} &= \mathcal{L}_S \left[ \delta_{ij} + \ell_1 E_i E_j + \ell_2 B_i B_j + \ell_3 (E_i B_j + B_i E_j) \right],
        \\
        P_{ij} &= \mathcal{L}_S \left[ \delta_{ij} - \ell_1 B_i B_j - \ell_2 E_i E_j + \ell_3 (E_i B_j + B_i E_j) \right].
    \end{align}
\end{subequations}
In deriving the continuity equation \eqref{eq:cont}, we have made use of equalities such as $\nabla\cdot(\ve\times\vb)=\vb\cdot(\nabla\times\ve)-\ve\cdot(\nabla\times\vb)$ and the first equation of \eqref{eq:fieldeqA}.

The weak energy condition requires the energy density \eqref{eq:eng_den} to be always positive unless the plane wave vanishes. Therefore, $K_{ij}$ and $P_{ij}$ should be positive-definite. That is to say, all eigenvalues of $K_{ij}$ and $P_{ij}$ should be positive. By calculating their eigenvalues, we find the following conditions
\begin{align}\label{eq:wec}
    \mathcal{L}_S > 0,
    \nonumber\\
    1 + \frac{\ell_1}{2} \vE^2 + \frac{\ell_2}{2} \vB^2 + \ell_3 \vB\cdot\vE > 0,
    \nonumber\\
    1 - \frac{\ell_1}{2} \vB^2 - \frac{\ell_2}{2} \vE^2 + \ell_3 \vB\cdot\vE > 0,
    \nonumber\\
    1 + \ell_1 \vE^2 + \ell_2 \vB^2 + 2 \ell_3  \vE\cdot\vB + \left(\ell_1\ell_2 - \ell_3^2\right) (\vE\times \vB)^2 > 0,
    \nonumber\\
    1 - \ell_1 \vB^2 - \ell_2 \vE^2 + 2 \ell_3 \vE\cdot\vB + \left(\ell_1\ell_2 - \ell_3^2\right) (\vE\times \vB)^2 > 0,
    \nonumber\\
    \left(\ell_1 \vE^2 + \ell_2 \vB^2 + 2 \ell_3\vB\cdot\vE\right)^2 - 4 \left(\ell_1\ell_2 - \ell_3^2\right)(\vE\times \vB)^2 \geqslant 0,
    \nonumber\\
    \left(\ell_1 \vB^2 + \ell_2 \vE^2 - 2 \ell_3\vB\cdot\vE\right)^2 - 4 \left(\ell_1\ell_2 - \ell_3^2\right)(\vE\times \vB)^2 \geqslant 0.
\end{align}
These conditions will be helpful for deriving dispersion relations in the coming section.

The energy flux density \eqref{eq:eng_flu} generalizes the Poynting vector of plane waves in Maxwell electrodynamics to its counterpart in nonlinear electrodynamics in the presence of a constant uniform electric and magnetic background. Along the direction of the energy flux density, the plane wave propagates, or equivalently, the light ray travels. As can be inferred from \eqref{eq:eng_flu}, this direction is perpendicular to the electric field $\ve$ of the plane wave.

\section{Dispersion relations and velocities}\label{sec:DR}
\subsection{Combined electric and magnetic background fields}\label{subsec:Comb}
In this subsection, to make our investigation slightly general, we assume the background is given by combined electric and magnetic fields. We are interested in plane wave solutions, $\ve=\ve_0\e^{\I(\vk\cdot\vx-\omega t)}$ and $\vb=\vb_0\e^{\I(\vk\cdot\vx-\omega t)}$\footnote{Or equivalently, $\ve=\ve_0\sin(\vk\cdot\vx-\omega t)$ and $\vb=\vb_0\sin(\vk\cdot\vx-\omega t)$.} with nonzero frequency $\omega$. Substituting them into \eqref{eq:fieldeqA}, we find the relation $\vb = (\vk/\omega)\times\ve$ which indicates that the magnetic component of the plane wave is always perpendicular to the wave vector. Substitution of them into \eqref{eq:fieldeqB} yields $\vn\cdot\ve=0$ with
\begin{align}\label{eq:n}
    \vn &= \vk
    +\ell_1 \vk\cdot\vE \left(\vE + \frac{\vk\times\vB}{\omega}\right)
    +\ell_2 \vk\cdot\vB \left(\vB - \frac{\vk\times\vE}{\omega}\right)
    +\ell_3 \vk\cdot\vB \left(\vE + \frac{\vk\times\vB}{\omega}\right)
    +\ell_3 \vk\cdot\vE \left(\vB - \frac{\vk\times\vE}{\omega}\right).
\end{align}
It is noteworthy that the three vectors $\vS$, $\vb$ and $\vn$ are not linearly independent, because they are all perpendicular to $\ve$. In contrast, unlike in Maxwell electrodynamics, the wave vector $\vk$ is not always perpendicular to $\ve$ here.

At the same time, if one combines the plane wave solutions and the relation $\vb = (\vk/\omega)\times\ve$ with \eqref{eq:fieldeqC} and then eliminates $\vb_0$, one will get linear equations of $\ve_0$ which can be arranged in a matrix form $M_{ij}e_{0j} = 0$, where $e_{0j}$ are components of the polarized electric field amplitude $\ve_0$ of the plane wave, and the coefficient matrix has the components of the form
\begin{align}
    M_{ij} &=
    \left( \omega^2 - \vk^2 \right) \delta_{ij}
    + k_i k_j
    + \ell_1 v_i v_j + \ell_2 u_i u_j - \ell_3 (u_iv_j
    + u_jv_i)
    \nonumber\\&\phantom{=}
    + \omega[ -\ell_1 (E_iv_j + E_jv_i) + \ell_2 (B_iu_j + B_ju_i)
    + \ell_3(E_iu_j + E_ju_i - B_iv_j - B_jv_i )]
    \nonumber\\&\phantom{=}
    +\omega^2 [\ell_1 E_iE_j + \ell_2 B_iB_j + \ell_3  (E_iB_j + E_jB_i)].
\end{align}
In the above expression, we have employed the notations $\bm{u} = \vE \times \vk$, $\bm{v} = \vB \times \vk$ for brevity, which will not be used anymore.

The system of homogeneous linear equations  $M_{ij}e_{0j} = 0$ yields nonvanishing solutions for $\ve$ if and only if its coefficient matrix is singular. That means the coefficient matrix has the zero determinant, $\det(M_{ij}) = 0$, which is a polynomial equation of $\omega^2$. After tedious but straightforward computations, we finally get four solutions or namely dispersion relations, $\omega=\omega_{1+}$, $\omega=\omega_{1-}$, $\omega=\omega_{2+}$ and $\omega=\omega_{2-}$ with
\begin{subequations}\label{eq:DR}
\begin{align}
    \omega_{1\pm} &= \pm \sqrt{
        \vk^2
        - d_1 \left[(\vB\times \vk)^2+(\vE\times \vk)^2\right]
        + d_1^2 (\vk\cdot\vE\times \vB)^2
        }
        + d_1 \vk\cdot\vE\times \vB,
    \\
    \omega_{2\pm} &= \pm \sqrt{
        \vk^2
        - d_2 \left[(\vB\times \vk)^2+(\vE\times \vk)^2\right]
        + d_2^2 (\vk\cdot\vE\times \vB)^2
        }
        + d_2 \vk\cdot\vE\times \vB,
\end{align}
\end{subequations}
which are distinguished by their subscripts. Hereafter we utilize some notations defined by
\begin{align}\label{eq:dW}
    d_{1} &= \frac{W_1 - W_2}{2W_0},    \quad
    d_{2} = \frac{W_1 + W_2}{2W_0},
    \nonumber\\
    W_0 &= 1 + \ell_1 \vE^2 + \ell_2 \vB^2 + 2 \ell_3 \vE\cdot\vB
    + (\ell_1\ell_2 - \ell_3^2) (\vE\times \vB)^2 ,
    \nonumber\\
    W_1 &= \ell_1 + \ell_2 + (\ell_1\ell_2 - \ell_3^2)(\vE^2 + \vB^2),
    \nonumber\\
    W_2 ^2
    &= \left[ \ell_1-\ell_2 - (\ell_1\ell_2 - \ell_3^2)\left(\vE^2-\vB^2\right) \right]^2
    +4\left[ \ell_3 - (\ell_1\ell_2 - \ell_3^2) \vB\cdot\vE \right]^2.
\end{align}
When deriving these dispersion relations, we have used some conditions inferred from \eqref{eq:wec}, e.g., $W_0>0$.

Before proceeding, it is crucial to take a closer look at dispersion relations \eqref{eq:DR}. By definition, the frequency of a plane wave is positive. But this is not guaranteed in \eqref{eq:DR}. For theories whose Lagrangian $\mathcal{L}(S,P)$ satisfies
\begin{equation}\label{eq:fc}
    2W_0 - W_1(\vE^2 + \vB^2) \geq |W_2| (\vE^2 + \vB^2),
\end{equation}
we have $\vk^2 - d_i \left[(\vB\times \vk)^2+(\vE\times \vk)^2\right]\geq0$ and thus $\omega_{i+}\geq0$, $\omega_{i-}\leq0$ for both $i=1$ and $i=2$, irrespective of the direction of the wave vector $\vk$. By contrast, when condition \eqref{eq:fc} is violated, the signs of $\omega_{i\pm}$ are sensitive to the direction of the wave vector.\footnote{The authors are grateful to the anonymous referee for valuable comments on dispersion relations, which were important for improving the manuscript.}

It is unlikely to prove condition \eqref{eq:fc} from \eqref{eq:wec}. From now on, we restrict ourselves to theories subject to this condition, focus on plane waves with dispersion relations $\omega=\omega_{1+}$ and $\omega=\omega_{2+}$ and omit the subscript $+$. For later convenience, we reexpress $\omega_{1+}$ and $\omega_{2+}$ in \eqref{eq:DR} as
\begin{equation}\label{dr}
    \omega_i^2 = \vk^2\left( 1 - a_i \right)    ,\quad
    i = 1, 2,
\end{equation}
where $a_1$ and $a_2$ are corrections from nonlinear electrodynamics,
\begin{align}\label{eq:DR12}
    a_i &= d_i \left[ (\vB\times \hvk)^2 + (\vE\times \hvk)^2 \right]
        -2 d_i^2 (\hvk\cdot\vE\times \vB)^2
    - 2d_i \hvk\cdot\vE\times \vB
    \sqrt{
            1
            - d_i [(\vB\times \hvk)^2+(\vE\times \hvk)^2]
            + d_i^2 (\hvk\cdot\vE\times \vB)^2
        }
\end{align}
with $\hvk$ being the unit wave vector $\vk/|\vk|$.

With the obtained dispersion relations, we can calculate the phase velocities and the refractive indices,
\begin{equation}
    v_{pi} = \frac{\omega_{i}}{|\vk|} = n_{i}^{-1} = \sqrt{1 - a_{i}},
\end{equation}
as well as the group velocities
\begin{widetext}
\begin{align}\label{eq:group_velocity}
    \vv_{gi}  = \frac{\d \omega_{i}}{\d \vk}
    = \frac{
        \hvk
        + d_i \left[ \vB\times(\vB\times\hvk) + \vE\times(\vE\times\hvk) \right]
        + d_i^2 (\hvk\cdot\vE\times\vB) (\vE\times\vB)
    }{\sqrt{1-d_i (\vB\times \hvk)^2 - d_i (\vE\times \hvk)^2 + d_i^2 (\hvk\cdot\vE\times \vB)^2}}
    + d_i \vE\times \vB.
\end{align}
\end{widetext}

Analogous to the optical birefringence in crystals, when the electromagnetic plane wave enters a region of uniform electromagnetic fields, the birefringence is expected to appear in nonlinear electrodynamics. It is dubbed as vacuum birefringence in the literature. There are two exceptions. First, when $(\vB\times \hvk)^2+(\vE\times \hvk)^2=2\hvk\cdot\vE\times \vB$, we find $a_1=a_2=0$ and thus vacuum birefringence disappears. Such a condition can be satisfied if $\vB$, $\vE$, $\hvk$ are mutually orthogonal and $|\vB|=|\vE|$ at the same time. Second, for theories satisfying $W_2 = 0$, that is, $2S (\ell_1\ell_2 - \ell_3^2) = \ell_1-\ell_2$ and $P (\ell_1\ell_2 - \ell_3^2) = \ell_3$, it is easy to see $d_1=d_2$ and thus $a_1=a_2$. There is no vacuum birefringence in such theories, as has been studied previously in Refs. \cite{Bialynicki-Birula1983, Russo_2023}.

\subsection{Pure electric or magnetic background field}\label{subsec:Pure}
Theoretically, the general dispersion relations \eqref{eq:DR} can be simplified a lot in special situations. Experimentally, the most accessible cases are the one with a pure electric external field and the one with a pure magnetic external field. This subsection will be dedicated to these cases. Recall that the powers of $P$ are even in Lagrangian $\mathcal{L}(S,P)$, then it is easy to see $\ell_3=0$ in both cases.

In the case of a pure electric field, we have
\begin{subequations}
\begin{align}
    a_{1} &= d_1(\vE\times \hat{\vk})^2 = \frac{\ell_1 (\vE\times \hat{\vk})^2 }{1 + \ell_1 \vE^2},
    \\
    a_{2} &= d_2(\vE\times \hat{\vk})^2 = \ell_2 (\vE\times \hat{\vk})^2 .
\end{align}
\end{subequations}
Likewise, in the case of a pure magnetic field, we find
\begin{subequations}
\begin{align}
    a_{1} &= d_1(\vB\times \hat{\vk})^2 = \frac{\ell_2 (\vB\times \hat{\vk})^2 }{1 + \ell_2 \vB^2},
    \\
    a_{2} &= d_2(\vB\times \hat{\vk})^2 = \ell_1 (\vB\times \hat{\vk})^2.
\end{align}
\end{subequations}
Obviously they are of the similar forms. Therefore, in the following investigation, we will only report the case of a pure magnetic field, while the case of a pure electric field can be obtained by directly replacing $\vB$ with $\vE$ and exchanging $\ell_1$ and $\ell_2$.

In the case of a pure magnetic background field, the phase velocities $\vv_{pi}$ and the group velocities $\vv_{gi}$ get simplified significantly \cite{PhysRevD.104.015006},
\begin{subequations}
\begin{align}
    \vv_{p1} &= \sqrt{1 - a_1} ~\hvk = \sqrt{ \frac{1 + \ell_2 (\vB\cdot\hvk)^2}{1 + \ell_2 \vB^2} } ~\hvk,
    \\
    \vv_{p2} &= \sqrt{1 - a_2} ~\hvk = \sqrt{1 - \ell_1 (\vB\times \hat{\vk})^2} ~\hvk,
    \\
    \vv_{g1} &= \frac{\d \omega_1}{\d \vk}
    = \frac{\hvk + \ell_2 \vB(\vB\cdot\hvk)}{ \sqrt{ 1 + \ell_2 \vB^2 }\sqrt{1 + \ell_2 (\vB\cdot\hvk)^2} },
    \\
    \vv_{g2} &= \frac{\d \omega_2}{\d \vk}
    = \frac{\hvk + \ell_1 \vB\times(\vB\times\hvk)}{\sqrt{1 - \ell_1 (\vB\times\hvk)^2}}.
\end{align}
\end{subequations}
In the limit of zero background field, $|\vB|\rightarrow 0$, it is easy to check that all phase velocities and group velocities go back to the Maxwell limit, $\vv_p = \vv_g = \hvk$.

\begin{figure}[htbp]
    \centering
    \includegraphics[width=0.45\textwidth]{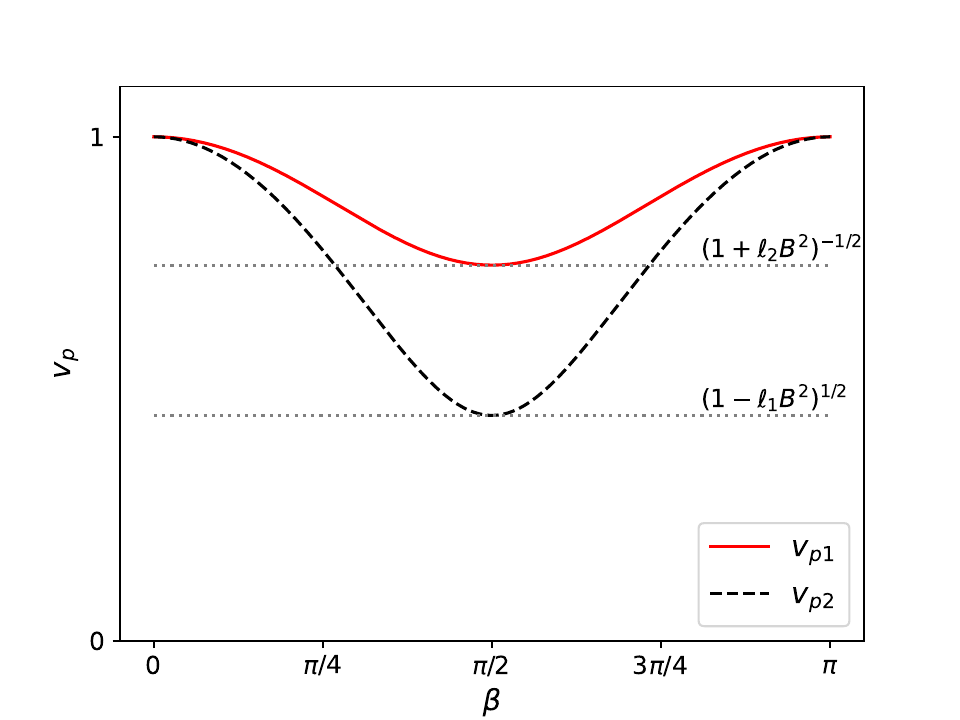}
    \includegraphics[width=0.45\textwidth]{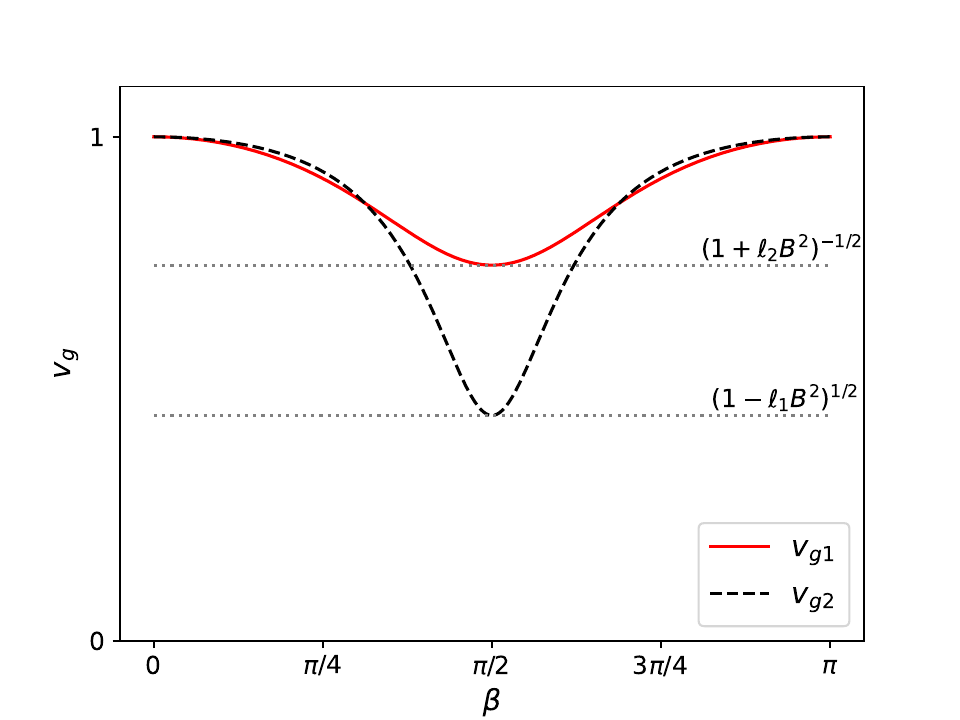}
    \caption{Illustration of phase velocities (left panel) and group velocities (right panel) as functions of $\beta$, i.e., the angle between $\vB$ and $\vk$.
    }\label{fig:vpvg}
\end{figure}

It is enlightening to examine how the velocities are changed by varying the angle $\beta$ between the magnetic background field $\vB$ and the wave vector $\vk$. On the one hand, the Maxwell limit $\vv_p = \vv_g = \hvk$ can be achieved when the wave vector is parallel or antiparallel to the background field, $\beta=0$ or $\pi$. On the other hand, when the wave vector is perpendicular to the background field, $\beta=\pi/2$, the corrections of nonlinear electrodynamics are maximal to phase velocities and group velocities, which become $\vv_{p1} = \vv_{g1} = (1 + \ell_2 \vB^2)^{-1/2}\hvk$, $\vv_{p2} = \vv_{g2} = (1 - \ell_1 \vB^2)^{1/2}\hvk$. The influences of general values of $\beta$ on phase velocities and on group velocities are illustrated in Figure \ref{fig:vpvg}. To avoid superluminosity, we assume that $\ell_1\geqslant 0$, $\ell_2\geqslant 0$ in this subsection.

\begin{figure}[htbp]
    \centering
    \includegraphics[width=0.45\textwidth]{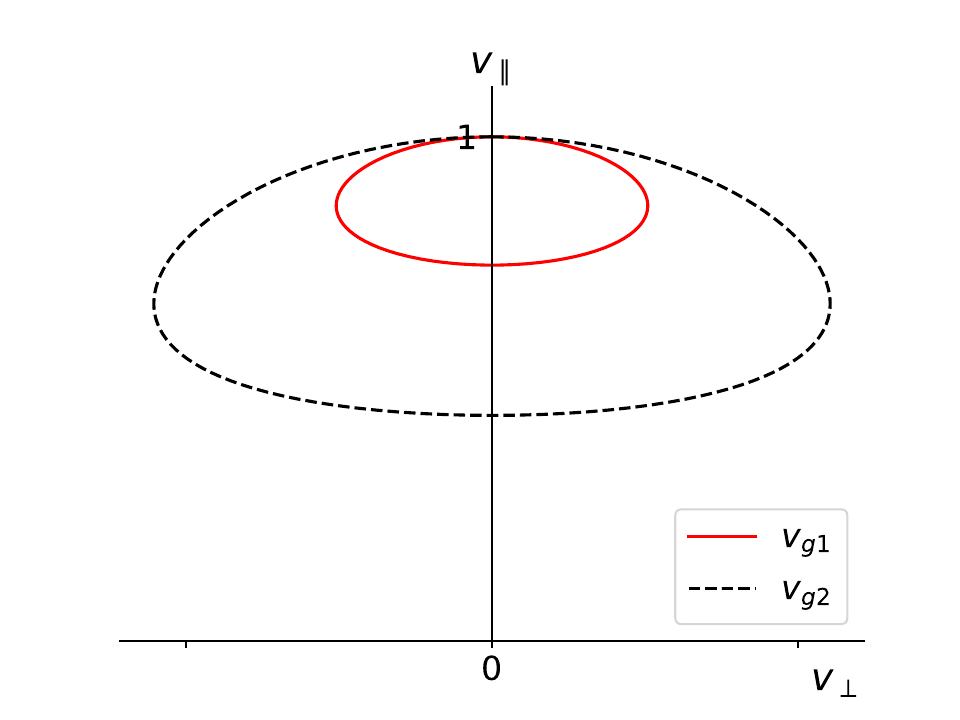}
    \caption{Trajectories of the heads of group velocity vectors by varying $\beta$ while fixing $|\vB|$ and $\vk$. All the tails of group velocity vectors are fixed to the origin. The coordinates $v_{\parallel}$ and $v_{\perp}$ denote the velocity components parallel and perpendicular to $\vk$, respectively.
    }\label{fig:v_group}
\end{figure}

Alternatively, we can also animate the absolute value of group velocities and their relative directions with respect to phase velocities. The latter always point in the direction of the wave vector. For this purpose, we fix the direction of wave vector $\vk$ and the value of $|\vB|$, and then change the direction of $\vB$ continuously in a plane. The resulting trajectories of $\vv_{g1}$ and $\vv_{g2}$ in this plane are animated in Figure \ref{fig:v_group}, where $v_{\parallel}$ and $v_{\perp}$ axes are directions parallel and perpendicular to $\vk$, respectively. Analytical calculation shows that
\begin{equation}
    \frac{|v_{g1\perp}|}{v_{g1\parallel}} \leqslant \frac{\ell_2\vB^2}{2 \sqrt{1 + \ell_2\vB^2}}
    ,\quad
    \frac{|v_{g2\perp}|}{v_{g2\parallel}} \leqslant \frac{\ell_1\vB^2}{2 \sqrt{1 - \ell_1\vB^2}}
\end{equation}
in good agreement with Figure \ref{fig:v_group}.

\section{Compton scattering}\label{sec:scatt}
The Compton scattering is the best known experiment that corroborates the particle nature of electromagnetic radiations. The dispersion relation of plane waves, interpreted as the energy-momentum relation of photons, plays a key role in explaining the Compton effect. Therefore, it is expected that this process can put constraints on nonlinear theories of electrodynamics. In the previous section, we have recast the dispersion relations of nonlinear electrodynamics in the unified form \eqref{dr}. This is very convenient for our study of the Compton effect.

In an electromagnetic background, consider an incoming photon of four-momentum ($\omega$, $\vk$) colliding with a free electron of mass $m_e$ in atoms at rest. At the instant of collision, the external force from the background field is negligible, therefore conservation of energy and momentum requires
\begin{equation}\label{eq:conservation}
    \omega + m_e = \omega' + E_e' ,\quad
    \vk = \vk' + \vp_e',
\end{equation}
in which ($\omega'$, $\vk'$) is the four-momentum of the outgoing photon, while ($E_e'$, $\vp_e'$) is the kinetic four-momentum of the outgoing electron immediately after the collision. Substituting them into the energy-momentum relation of the electron, $E_e'^{2}=\vp_e'^{2}+m_e^2$, we get
\begin{equation}\label{eq:collision}
    \omega^2 - \vk^2 + \omega'^2 - \vk'^2 - 2(\omega\omega'-|\vk||\vk'|\cos\theta) = 2m_e(\omega'-\omega),
\end{equation}
where $\theta$ is the deflection angle of the wave vector so that $\vk\cdot\vk'=|\vk||\vk'|\cos\theta$.

In our notations, the Compton wavelength of an electron is $\lambda_e = 2\pi/m_e$, while the wavelength of a photon is $\lambda = 2\pi/|\vk|$. In accordance with \eqref{dr}, into \eqref{eq:collision} we will insert the dispersion relations
\begin{equation}
    \omega^2 = \vk^2\left( 1 - a \right) ,\quad
    \omega'^2 = \vk'^2\left( 1 - a' \right)
\end{equation}
for nonlinear electrodynamics rather than the standard dispersion relation for Maxwell electrodynamics. After doing this and defining $\lambda' = 2\pi/|\vk'|$, we arrive at
\begin{align}
    &\sqrt{1-a} \lambda'
    - \sqrt{1-a'} \lambda
    = 2 \lambda_e\sin^2\frac{\theta}{2}
    +\frac{1}{2} \lambda_e \left(
    \frac{a \lambda '}{\lambda }
    +\frac{a' \lambda}{\lambda'}
    +2 \sqrt{1-a} \sqrt{1-a'} - 2
    \right).
\end{align}
Consequently, the wavelength after scattering is given by
\begin{widetext}
\begin{align}\label{eq:Delta}
    \frac{\lambda' }{\lambda}
    &= \frac{
        2 \lambda_e \sin^2\frac{\theta}{2}
        + \sqrt{1-a'} \lambda
        + (\sqrt{1-a} \sqrt{1-a'} - 1) \lambda_e
    }{ 2 \sqrt{1-a} \lambda - a \lambda_e}
    \nonumber\\&\phantom{=}
    \pm \frac{ \sqrt{
        \left[
            2 \lambda_e \sin^2\frac{\theta}{2}
            + \sqrt{1-a'} \lambda
            +(\sqrt{1-a} \sqrt{1-a'} - 1) \lambda_e
        \right]^2
        + a' \lambda_e \left(2 \lambda  \sqrt{1-a} - a \lambda_e\right)
    } }{ 2 \sqrt{1-a} \lambda - a \lambda_e}.
\end{align}
\end{widetext}
Hereafter we use a prime to highlight quantities for outgoing photons.

The form of \eqref{eq:Delta} is very intricate, but we can make it simpler in two special circumstances.

On condition that $|a|+|a'|$ is much less than both $1$ and $\lambda/\lambda_e$, we can expand the wavelength shift $\Delta\lambda=\lambda'-\lambda$ to the first order of $a$ and $a'$,
\begin{subequations}
\begin{align}
    \Delta\lambda_+ &\approx 2 \lambda_e \sin^2\frac{\theta}{2}
    + a \left[
        \frac{\lambda }{2}
        + \frac{\lambda_e \left(\lambda_e+\lambda \right)}{\lambda} \sin^2\frac{\theta}{2}
    \right]
    -a' \left(
        \frac{\lambda }{2}
        +\frac{\lambda_e^2 \sin^2\frac{\theta}{2}}{2 \lambda_e \sin^2\frac{\theta}{2}+ \lambda }
    \right),
    \label{eq:DeltaA}\\
    \Delta\lambda_- &\approx
    -\frac{a' \lambda  \lambda_e}{4 \lambda_e \sin^2\frac{\theta}{2}+2\lambda }
    -\lambda.
    \label{eq:DeltaB}
\end{align}
\end{subequations}
The solution \eqref{eq:DeltaB} should be discarded because it has unphysical implications. If $a'>0$, it yields $\lambda'<0$. When $0\leqslant-a'\ll1$, it leads to $\omega'\gg\omega$ in contradiction with \eqref{eq:conservation}. In the Thomson limit, $\lambda \gg \lambda_e$, the wavelength shift \eqref{eq:DeltaA} is approximately
\begin{equation}
    \Delta\lambda \approx
    \lambda_e \sin^2\frac{\theta}{2}\left( 2 + a \right)
    + \frac{\lambda }{2} \left( a - a' \right).
\end{equation}

In the circumstance that $|a|+|a'|\ll1$ and $\lambda/\lambda_e\ll\sin ^2(\theta/2)$, we can expand the wavelength shift $\Delta\lambda=\lambda'-\lambda$ to the leading order of $a$, $a'$ and $\lambda/\lambda_e$,
\begin{subequations}
\begin{align}
    \Delta\lambda_+ &\approx
    \left[\frac{4 \lambda  \lambda _e }{2 \lambda -a \lambda _e}
    +\frac{4a \lambda^2  \lambda _e }{\left(2 \lambda -a \lambda _e \right){}^2}\right] \sin^2\frac{\theta}{2}
    -\frac{a' \lambda  \lambda _e}{2 \lambda -a \lambda _e},
    \label{eq:DeltaC}\\
    \Delta\lambda_- &\approx
    -\lambda \left( 1 + \frac{a'}{4\sin^2\theta} \right).
    \label{eq:DeltaD}
\end{align}
\end{subequations}
Note the solution \eqref{eq:DeltaC} is dominated by the first term, which is much larger than $\lambda$ in this circumstance. To guarantee $\lambda'>0$ in this solution, it is necessary that $\lambda/\lambda_e>a/2$. \eqref{eq:DeltaD} is physically acceptable only if $-a'/(4\sin^2\theta)>\sqrt{(1-a')/(1-a)}$, otherwise it would violate the condition $\omega'\leqslant\omega$ imposed by \eqref{eq:conservation}.

\section{Applications to BLST electrodynamics}\label{sec:inBLST}
Armed with the above results, in this section let us apply them to the BLST electrodynamics whose Lagrangian can be written as \cite{Bandos_2021,Sorokin_2022}
\begin{equation}\label{LBLST}
    \mathcal{L}_\mathrm{BLST}
    = T-\sqrt{T^2-2T\mathcal{L}_\mathrm{ModMax}-P^2},
\end{equation}
where the Lagrangian of ModMax electrodynamics\cite{Bandos_2020}
\begin{equation}\label{LModMax}
    \mathcal{L}_\mathrm{ModMax}=\cosh\gamma S + \sinh\gamma\sqrt{S^2+P^2}.
\end{equation}
Here $T$ and $\gamma$ are nonnegative constant parameters. In the literature, they are usually referred to as Born-Infeld constant and ModMax constant, respectively. As explained in Refs. \cite{Bandos_2021,ferko2023dualityinvariant,shi2023transverse}, Maxwell, Born-Infeld, Bialynicki-Birula and ModMax theories can be recovered from BLST electrodynamics in appropriate limits. Note that $\gamma$ is dimensionless, whereas $T$ has the dimension of $S$ or $P$ (or equivalently, $\vE^2$ or $\vB^2$).

In the current section, we will not consider the appearance of an external electric field. This is not only for simplicity, but also for practical reasons. Practically, the electron cannot stay at rest in an external electric field unless it is balanced out by the internal electric field of atoms.

\subsection{Dispersion relations}\label{subsec:DRinBLST}
As can be seen from \eqref{eq:dW}, the dispersion relations depend on $\ell_1$, $\ell_2$ and $\ell_3$. Their full expressions are lengthy for BLST electrodynamics,
\begin{widetext}
\begin{subequations}
\begin{align}
    \ell_1
    &= \frac{T \cosh (2 \gamma ) \left(P^2-2 S^2\right) \sqrt{P^2+S^2}-3 P^2 T \sqrt{P^2+S^2}+2 \sinh \gamma  \left(P^4-P^2 T^2-2 S^3 T \cosh \gamma \right)}{2 \left(P^2+S^2\right) \left(\cosh \gamma  \sqrt{P^2+S^2}+S \sinh \gamma \right) \left(2 T \sinh \gamma  \sqrt{P^2+S^2}+P^2+2 S T \cosh \gamma -T^2\right)},    
    \\
    \ell_2
    &= -\ell_1 +
    \frac{2 S \cosh \gamma  \sqrt{P^2+S^2}+\sinh \gamma  \left(P^2+2 S^2-T^2\right)-2 T \sqrt{P^2+S^2}}{\left(\cosh \gamma  \sqrt{P^2+S^2}+S \sinh \gamma \right) \left(2 T \sinh \gamma  \sqrt{P^2+S^2}+P^2+2 S T \cosh \gamma -T^2\right)},
    \\
    \ell_3
    &= -\frac{P \left\{S \sinh \gamma  \left(3 T \sinh \gamma  \sqrt{P^2+S^2}+2 P^2+S^2-T^2\right)+\cosh \gamma  \left[T \sinh \gamma  \left(P^2+3 S^2\right)+\left(P^2+S^2\right)^{3/2}\right]\right\}}{\left(P^2+S^2\right) \left(\cosh \gamma  \sqrt{P^2+S^2}+S \sinh \gamma \right) \left(2 T \sinh \gamma  \sqrt{P^2+S^2}+P^2+2 S T \cosh \gamma -T^2\right)}.    
\end{align}
\end{subequations}
\end{widetext}
However, when the external electric field is turned off, one has simply $S=-\vB^2/2$, $P=0$ and hence
\begin{align}
    \mathcal{L}_S = \frac{\e^{-\gamma } T}{\sqrt{T \left(\vB^2 \e^{-\gamma }+T\right)}} ,\quad
    &\ell_1 = \frac{1}{\vB^2+\e^{\gamma } T} ,
    \quad
    \ell_2 
    = \e^{\gamma } \left(\frac{2 \sinh \gamma }{\vB^2}+\frac{1}{T}\right),\quad
    \ell_3 = 0.
\end{align}

In this case, the results of Subsection \ref{subsec:Pure} are applicable, yielding the dispersion relations
\begin{subequations}\label{eq:dispBLST}
\begin{align}
    \omega^2 &= \omega_1^2 = \vk^2 \left[1 - \frac{(1 - e^{-2\gamma} +\tvB^2 )\sin^2 \beta }{1+\tvB^2}\right],
    \\
    \omega^2 &= \omega_2^2 = \vk^2 \left(1 - \frac{\tvB^2 \sin^2 \beta }{1+\tvB^2}\right).
\end{align}
\end{subequations}
Here we have introduced the notations $\tvB^2=\vB^2/(\e^{\gamma}T)$ and $\hvB=\vB/|\vB|$. In the limit $T\rightarrow\infty$, they reduce to equations (6.1) and (6.2) in Ref. \cite{Lechner_2022} for ModMax electrodynamics. We warn that the limit $\vB\rightarrow0$ is ill-defined because, by assumption in Section \ref{sec:framework}, the background field should be intense in comparison with the plane wave. It is also noteworthy that $\omega_2$ is simply dictated by $\tvB$.

The dispersion relations \eqref{eq:dispBLST} indicate there are two types of plane waves in a constant uniform magnetic background in BLST electrodynamics. Henceforth they will be dubbed as type 1 wave and type 2 wave for clarity. Their phase velocities and group velocities are correspondingly
\begin{subequations}
\begin{align}
    \vv_{p1} &= \sqrt{1 - \frac{(1 - \e^{-2\gamma} +\tvB^2 )\sin^2 \beta }{1+\tvB^2}} ~\hvk,
    \\
    \vv_{p2} &= \sqrt{1 - \frac{\tvB^2 \sin^2 \beta }{1+\tvB^2}} ~\hvk,\\
    \vv_{g1} &= \frac{\e^{-2\gamma} \hvk + \left(1-\e^{-2\gamma}+\tvB^2\right)\cos\beta\hvB}
    { \sqrt{1+\tvB^2}
    \sqrt{\e^{-2\gamma}\sin^2\beta + (1+\tvB^2)\cos^2\beta} },
    \\
    \vv_{g2} &= \frac{\hvk + \tvB^2\cos\beta\hvB}
    {\sqrt{1+\tvB^2}\sqrt{1+\tvB^2\cos^2\beta}}.
\end{align}
\end{subequations}
By our assumption $\gamma\geqslant0$, the type 1 wave has a slower phase velocity than the type 2 wave, and we have $n_1\geqslant n_2$. In addition, it is straightforward to prove that
\begin{subequations}
\begin{align}
    \vv_{g1}^2 &= 1 - \frac{\left(1-\e^{-2\gamma}+\tvB^2\right) \sin^2\beta}
    {\left(1+\tvB^2\right) \left[\sin^2\beta + \e^{2\gamma} (1+\tvB^2) \cos^2\beta\right]},
    \\
    \vv_{g2}^2 &= 1 - \frac{\tvB^2 \sin ^2\beta}
    {\left(1+\tvB^2\right) \left(1+\tvB^2\cos^2\beta\right)}.
\end{align}
\end{subequations}
Neither of the group velocities is faster than the light velocity in vacuum.

\subsection{Energy flux density}\label{subsec:fluxinBLST}
In applications, starting with \eqref{eq:n}, we can determine the direction the electric field $\ve$ of the plane wave according to $\vn\cdot\ve=0$, and then work out the direction of the energy flux density $\vS$ with the help of \eqref{eq:eng_flu} and $\vb = (\vk/\omega)\times\ve$.

For BLST electrodynamics in a constant uniform magnetic background $\vB$, \eqref{eq:n} takes the form
\begin{align}\label{eq:nBLST}
    \vn &= \vk
    +\e^{\gamma } \left(\frac{2 \sinh \gamma }{\vB^2}+\frac{1}{T}\right) (\vk\cdot\vB) \vB
    \nonumber\\
    &= k \hvk + k\e^{2\gamma}\left(1-\e^{-2\gamma}+\tvB^2\right)\cos\beta\hvB
\end{align}
which is parallel to $\vv_{g1}$, and \eqref{eq:eng_flu} becomes
\begin{align}\label{eq:SBLST}
    \vS &= \mathcal{L}_S \ve\times\left[
    \vb
    -\frac{\vB(\vB\cdot\vb)}{\vB^2+\e^{\gamma } T}\right]
    \nonumber\\
    &= \frac{\mathcal{L}_S k}{\omega} \left[
    \ve^2\hvk-(\hvk\cdot\ve)\ve
    -\frac{\tvB^2(\hvk\cdot\ve\times\hvB)}{1+\tvB^2}(\ve\times\hvB)
    \right].
\end{align}

If the wave vector $\vk$ is parallel or antiparallel to the background magnetic field $\vB$, then \eqref{eq:nBLST} tells us that $\vn$ is collinear to $\vB$. Recalling $\vn\cdot\ve=0$ and $\vb = (\vk/\omega)\times\ve$, we can infer that $\ve$ and $\vb$, the electric and magnetic components of the plane wave, are both orthogonal to $\vB$ in this case. In consequence, the energy flux density \eqref{eq:eng_flu} is reduced to $\vS=\mathcal{L}_S\ve\times\vb=\mathcal{L}_S\ve^2\vk/\omega$.

Otherwise, if $\vk$ and $\vB$ are two linearly independent vectors, then they can span a 2-dimensional linear space in which the vector $\vn$ lies, and $\hvk\times\hvB$ will be a well-defined unit vector perpendicular to $\vn$. Perpendicular to $\vn$, there is another unit vector in this 2-dimensional linear space. It can be written as $\hvm=\vm/|\vm|$ in terms of
\begin{subequations}
\begin{align}
    &\vm = \left(1+\tvB^2\right)\cot\beta\left(\hvk-\cos\beta\hvB\right) - \e^{-2\gamma}\sin\beta\hvB,
    \\
    &|\vm| = \sqrt{\e^{-4\gamma}\sin^2\beta+(1+\tvB^2)^2\cos^2\beta} = \frac{|\vn|}{k\e^{2\gamma}}.
\end{align}
\end{subequations}
In this case, the electric component of the plane wave is $\ve=\ve_0\e^{\I(\vk\cdot\vx-\omega t)}$ with $\ve_0 = |\ve_0|(\cos\chi\hvk\times\hvB + \sin\chi\hvm)$, where $\chi$ is the angle between $\ve$ and $\hvk\times\hvB$. Inserting it into \eqref{eq:SBLST}, we find
\begin{align}\label{eq:S0BLST}
    \vS &=\frac{\mathcal{L}_S k \ve^2}{\omega} \left\{
    \cos^2\chi \left[\hvk
    -\frac{\tvB^2\sin^2\beta}{1+\tvB^2} \left(\hvk-\cos\beta\hvB\right)\right]
    \right.
    \nonumber\\&\phantom{=}
    +\left[\e^{-2\gamma}\sin^2\beta+\left(1+\tvB^2\right)\cos^2\beta\right]
    \sin^2\chi \frac{\hvn}{|\vm|}
    \nonumber\\&\phantom{=}
    -\left(1-\e^{-2\gamma}\right)\sin\beta\cos\beta\sin\chi\cos\chi\frac{\hvk\times\hvB}{|\vm|}
    \Biggr\},
\end{align}
which can be parallel to $\vn$ and $\vv_{g1}$ if $\chi=\pi/2$.

In the limit $\beta=0$ or $\pi$, it is not hard to check that $\hvn=\hvk$ and $|\vm|=1+\tvB^2$. As a result, one can reproduce $\vS=\mathcal{L}_S\ve^2\vk/\omega$ by extrapolating \eqref{eq:S0BLST} to these limits.

The limit $\beta=\pi/2$ is also of particular interest. In this limit, the wave vector $\vk$ is perpendicular to the background magnetic field $\vB$, and it is provable that $\hvn=\hvk$, $\hvm=-\hvB$ and
\begin{align}
    \vS = \frac{\mathcal{L}_S \ve^2}{\omega}
    \left( \frac{\cos^2\chi}{1+\tvB^2} + \sin^2\chi \right)
    \vk
\end{align}
which attains its minimal amplitude when $\ve$ is perpendicular to $\vB$.

\subsection{Compton scattering}\label{subsec:scattinBLST}
\begin{figure}[htbp]
    \centering
    \includegraphics[width=0.35\textwidth]{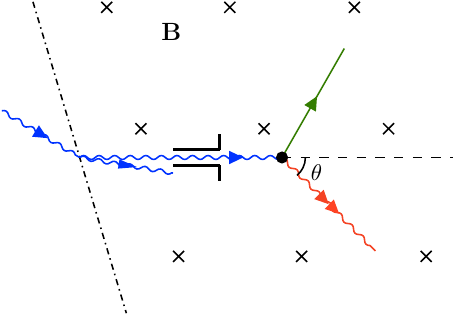}
    \caption{A schematic diagram of the designed experiment in Subsection \ref{subsec:scattinBLST}. With its boundary denoted by a dash-dotted line, the magnetic field is normal to the page. Accordingly, in the plane of the page, the unit vector $\hvl$ normal to the field boundary is perpendicular to the dash-dotted line. From left to right, the photon experiences birefringence and Compton scattering in sequence.
    }\label{fig:ex}
\end{figure}

As depicted by Figure \ref{fig:ex}, in BLST electrodynamics, when a beam of electromagnetic plane wave enters a region of uniform magnetic field, it will undergo birefringence and split into two beams. In this process, its frequency is unchanged, but its wavelength is changed according to dispersion relations \eqref{eq:dispBLST}. The direction of wave vector is changed according to Snell's law in vector form $\hvi\times\hvl=n_i\hvk_i\times\hvl$, where $\hvi$ and $\hvk_i$ are the unit directional vectors of the incident and type $i$ transmitted rays respectively, $\hvl$ is the unit vector normal to the boundary of field region, and $n_i=1/|\vv_{pi}|$ is the refractive index of the type $i$ wave.

If the region of magnetic field is large enough, we can select one beam to do the Compton scattering experiment inside this region and measure the shifts of wavelength of scattered rays at different deflection angles. The experimental results can be then used to put constraints on nonlinear electrodynamics such as BLST theory. Since deviations from Maxwell electrodynamics are expected to be very small, precise theoretical predictions are important for designing the experiment and analyzing the results.

There is a pitfall. By definition, the wave vector points in the direction of phase velocity. In BLST electrodynamics, as happens in the optical birefringence in crystals, the direction of phase velocity is not necessarily identical to the direction of wave propagation. The latter is in the direction of wave energy flux (light ray), see \eqref{eq:SBLST} and \eqref{eq:S0BLST}. More generally, in nonlinear electrodynamics, one should be cautious that the wave vector is not always along the direction of light ray in the presence of external fields. This makes it challenging to determine the direction of wave vector in experiments.

The exceptions are the cases with $\beta=0$, $\pi/2$ and $\pi$. As we have pointed out in Subsection \ref{subsec:fluxinBLST}, in these cases, the wave vector $\vk$ is in the same direction of the energy flux density $\vS$, i.e., the direction of ray.

This inspires us to consider a configuration illustrated by Figure \ref{fig:ex}, in which the background magnetic field $\vB$ is parallel to its boundary so that $\hvl\perp\hvB$. Orthogonal to the background field, there is a ray incident from the vacuum, $\hvi\perp\vB$. It is not hard to see that $\hvi\times\hvl\parallel\vB$ and hence $\hvk_i\times\hvl\parallel\hvB$ in conformity with Snell's law. It means the refracted rays are also orthogonal to the background field in this configuration. Subsequently, one of the refracted rays collides with an electron and is scattered in all directions, but we will restrict ourselves to scattered rays in the plane $\beta=\pi/2$ (the plane of the page in Figure \ref{fig:ex}) which thus travel along the directions of their wave vectors.

In the vacuum, the incident wave has the wavelength $2\pi/\omega$. As a result of birefringence, the refracted waves in the magnetic field have the same frequency $\omega$ but different wave vectors $\vk_1$, $\vk_2$. Their dispersion relations can be read off from \eqref{eq:dispBLST} by setting $\beta=\pi/2$,
\begin{align}\label{eq:DRBLSTin}
    \omega^2 = \frac{e^{-2\gamma} \vk_1^2}{1+\tvB^2} ,\quad
    \omega^2 = \frac{\vk_2^2}{1+\tvB^2}.
\end{align}

\begin{figure}[htpt!]
    \centering
    \includegraphics[width=0.4\textwidth]{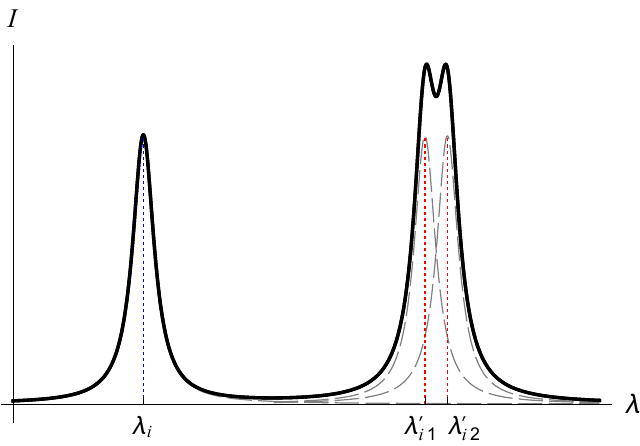}
    \caption{(not to scale). Illustration of the coherent peak and the modified Compton peaks in BLST electrodynamics. The horizontal axis represents the wavelength. The vertical axis represents the intensity of scattered rays, which is proportional to the number of scattered-photon counts in a certain direction. The coherent peak sits at $\lambda=\lambda_i$ (left blue dotted line), while the corresponding Compton peaks are located at $\lambda=\lambda'_{i1}$ and $\lambda=\lambda'_{i2}$ (right red dotted lines). As depicted by the black solid curve, the Compton peaks (green dashed curves) merge in low-precision experiments.
    }\label{fig:pk}
\end{figure}

After that, if the type $i$ wave is scattered off a free electron in atoms, then it is a Compton scattering process. Both the frequency and the wave vector are changed. Restricted to the plane $\beta=\pi/2$, the dispersion relations of scattered waves are
\begin{align}\label{eq:DRBLSTout}
    \omega'^2_{i1} = \frac{e^{-2\gamma} \vk'^2_{i1}}{1+\tvB^2} ,\quad
    \omega'^2_{i2} = \frac{\vk'^2_{i2}}{1+\tvB^2}.
\end{align}
Based on \eqref{eq:DeltaA}, the wavelengthes $\lambda'_{ij}=2\pi/|\vk'_{ij}|$ of scattered waves are related to $\lambda_i=2\pi/|\vk_i|$ by
\begin{align}
    \lambda'_{ij}-\lambda_i &\approx (2 + a_i) \lambda_e \sin^2\frac{\theta}{2}
    + a_i \left(
        \frac{\lambda_i }{2}
        + \frac{\lambda_e^2 \sin^2\frac{\theta}{2}}{\lambda_i}
    \right)
    -a_j \left(
        \frac{\lambda_i }{2}
        +\frac{\lambda_e^2 \sin^2\frac{\theta}{2}}{2 \lambda_e \sin^2\frac{\theta}{2}+ \lambda_i }
    \right)
\end{align}
with $\theta$ being the scattering angle, and
\begin{align}
    a_1 = \frac{1 - e^{-2\gamma} + \tvB^2 }{1+\tvB^2} ,\quad
    a_2 = \frac{\tvB^2 }{1+\tvB^2}.
\end{align}
The above expression of wavelength shift is valid under the condition that $\max\{a_1,a_2\}\ll\min\{1,\lambda/\lambda_e\}$, which can be met since $\gamma$ and $\tvB^2$ are small in reality.

Alternatively, if the type $i$ wave is scattered off a bound electron in atoms, then the coherent scattering occurs. During this process, neither the frequency nor the wavelength is changed in any scattering angle. As a result, there are outgoing waves with $\omega'_{i0}=\omega$, $\lambda'_{i0}=\lambda_i$ in all directions in the plane $\beta=\pi/2$.

\begin{figure}[htpt!]
    \centering
    \includegraphics[width=0.45\textwidth]{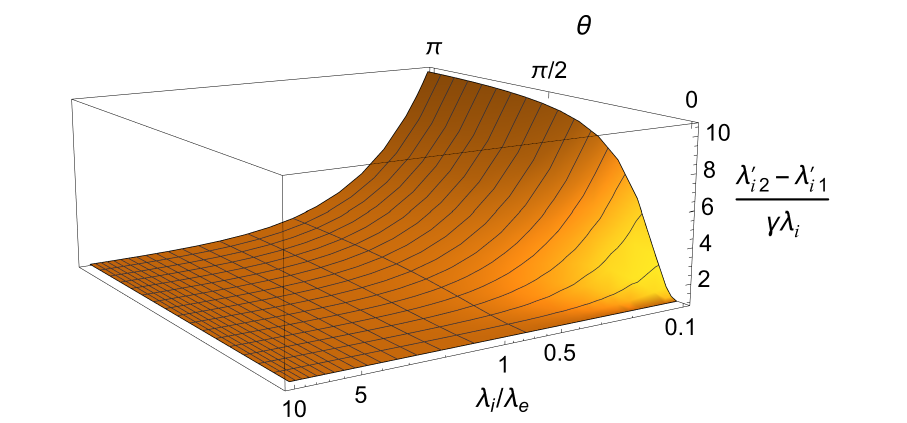}
    \caption{Illustration of $(\lambda'_{i2}-\lambda'_{i1})/(\gamma\lambda_i)$ in the contour $0\leqslant\theta\leqslant\pi$, $0.1\leqslant\lambda_i/\lambda_e\leqslant10$ according to \eqref{eq:wldCom}. It tends to $1$ in the small angle limit $\theta\ll1$ or in the long wavelength limit $\lambda_i/\lambda_e\gg1$, but tends to $\lambda_e/\lambda_i$ in the short wavelength limit $\lambda_i/\lambda_e\ll\sin ^2(\theta/2)$.
    }\label{fig:Delta}
\end{figure}

Theoretically, in the spectrum of outgoing photons there are three peaks, located at wavelengthes of $\lambda'_{i0}$, $\lambda'_{i1}$ and $\lambda'_{i2}$, respectively. We illustrate them in Figure \ref{fig:pk}. Experimentally, it should be viable to separate the coherent peak at a wavelength $\lambda'_{i0}$ with the modified Compton peaks at wavelengthes $\lambda'_{i1}$ and $\lambda'_{i2}$, but it will be challenging and important to distinguish the two Compton peaks, whose wavelength difference is
\begin{align}\label{eq:wldCom}
    \lambda'_{i2}-\lambda'_{i1} &\approx \gamma \left(
        \lambda_i
        +\frac{2 \lambda_e^2 \sin^2\frac{\theta}{2}}{2 \lambda_e \sin^2\frac{\theta}{2}+ \lambda_i }
    \right).
\end{align}
For short wavelength photons, $\gamma\ll\lambda_i/\lambda_e\ll\sin ^2(\theta/2)$, the difference can be approximated by $\lambda'_{i2}-\lambda'_{i1}\approx\gamma\lambda_e$ unless the scattering angle is small. In contrast, \eqref{eq:DRBLSTin} suggests that $\lambda_2-\lambda_1\approx2\pi\gamma/\omega\approx\gamma\lambda_i$. We can see the wavelength difference has been magnified by the Compton scattering in this situation. In the Thomson limit, $\lambda_i\gg\lambda_e$, the wavelength difference \eqref{eq:wldCom} is approximately $\lambda'_{i2}-\lambda'_{i1}\approx\gamma\lambda_i$. According to \eqref{eq:wldCom}, $(\lambda'_{i2}-\lambda'_{i1})/(\gamma\lambda_i)$  can be taken as a function of $\theta$ and $\lambda_i/\lambda_e$, which is depicted in the interval $\lambda_i/\lambda_e\in[0.1,10]$ in Figure \ref{fig:Delta}.

At last we turn to the wavelength difference between the coherent peak and one of the Compton peaks,
\begin{align}\label{eq:wldcoh}
    \lambda'_{ii}-\lambda_i &\approx (2 + a_i) \lambda_e \sin^2\frac{\theta}{2}
    + \frac{2 a_i \lambda_e^3 \sin ^4\left(\frac{\theta }{2}\right)}{\lambda_i\left(2 \lambda_e \sin^2\frac{\theta}{2}+ \lambda_i \right)}.
\end{align}
In the Thomson limit, $\lambda_i\gg\lambda_e$, it can be estimated by $\lambda'_{ii}-\lambda_i \approx (2 + a_i) \lambda_e \sin ^2(\theta/2)$. For short wavelength photons, $\gamma\ll\lambda_i/\lambda_e\ll\sin ^2(\theta/2)$, the modified Compton shift \eqref{eq:wldcoh} is estimatedly $\lambda'_{ii}-\lambda_i \approx (2 + a_i\lambda_e/\lambda_i) \lambda_e \sin ^2(\theta/2)$. Remembering that $a_1\approx2\gamma+\tvB^2$ and $a_2\approx\tvB^2$, we can see the Compton shift has been enhanced in this situation.


\section{Conclusion}\label{sec:conclusion}
In the presence of electric and magnetic background fields, after a general investigation of dispersion relations and Compton scattering of electromagnetic waves in nonlinear electrodynamics, we designed a specific Compton scattering experiment to test BLST electrodynamics. The modified Compton shift of wavelength has been worked out, which can be used to constrain parameters of BLST theory, $\gamma$ and $T$. Especially, in the situation $\gamma+\tvB^2\ll\lambda_i/\lambda_e\ll1$, we found
\begin{subequations}
\begin{align}
    \lambda'_{12}-\lambda'_{11} &= \lambda'_{22}-\lambda'_{21} \approx \gamma\lambda_e,
    \\
    \lambda'_{11}-\lambda_1 &\approx \left[2 + \frac{(\gamma+\tvB^2)\lambda_e}{\lambda_1}\right] \lambda_e \sin^2\frac{\theta}{2},
    \\
    \lambda'_{22}-\lambda_2 &\approx \left(2 + \frac{\tvB^2\lambda_e}{\lambda_2}\right) \lambda_e \sin^2\frac{\theta}{2}
\end{align}
\end{subequations}
barring small scattering angles, where $\tvB^2=\vB^2/(\e^{\gamma}T)$. Therefore, the two Compton peaks in Figure \ref{fig:pk} coincide theoretically if and only if $\gamma=0$ or, in other words, when BLST electrodynamics reduces to the Born-Infeld theory. As warned in Subsection \ref{subsec:DRinBLST}, we cannot set $\vB=0$ in the present article. Nevertheless, we can still take $T\rightarrow\infty$ to recover the limit of ModMax theory, in which there remain two Compton peaks. It is interesting to note that, in high-precision experiments, the fingerprints of finite $T$ and nonzero $\gamma$ can be searched for separately in the Compton shift $\lambda'_{22}-\lambda_2$ and the double-peak structure of Compton peak.

As cornerstones of atomic physics, experiments in external electric or magnetic fields, such as the (anomalous) Zeeman effect, the Stark effect and the Stern-Gerlach experiment, have given birth to new theories of physics which are not new today. In the future, it is expected that upcoming and planned experiments in intense background fields will reveal the path to new physics \cite{Fedotov_2023}. To the best of our knowledge, there are experiments to measure the vacuum magnetic birefringence \cite{Della_Valle_2014,EJLLI20201} but no experiment of magnetic Compton scattering so far. We hope there will be Compton scattering experiments in strong magnetic fields in the future, which are important tests not only for quantum electrodynamics but also for nonlinear theories of classical electrodynamics.

\acknowledgments{This work is sponsored by the Natural Science Foundation of Shanghai (Grant No. 24ZR1419300).}

\appendix
\pdfstringdefDisableCommands{%
\renewcommand*{\bm}[1]{#1}%
} 

\bibliography{ref}

\begin{thebibliography}{30}%
\makeatletter
\providecommand \@ifxundefined [1]{%
 \@ifx{#1\undefined}
}%
\providecommand \@ifnum [1]{%
 \ifnum #1\expandafter \@firstoftwo
 \else \expandafter \@secondoftwo
 \fi
}%
\providecommand \@ifx [1]{%
 \ifx #1\expandafter \@firstoftwo
 \else \expandafter \@secondoftwo
 \fi
}%
\providecommand \natexlab [1]{#1}%
\providecommand \enquote  [1]{``#1''}%
\providecommand \bibnamefont  [1]{#1}%
\providecommand \bibfnamefont [1]{#1}%
\providecommand \citenamefont [1]{#1}%
\providecommand \href@noop [0]{\@secondoftwo}%
\providecommand \href [0]{\begingroup \@sanitize@url \@href}%
\providecommand \@href[1]{\@@startlink{#1}\@@href}%
\providecommand \@@href[1]{\endgroup#1\@@endlink}%
\providecommand \@sanitize@url [0]{\catcode `\\12\catcode `\$12\catcode
  `\&12\catcode `\#12\catcode `\^12\catcode `\_12\catcode `\%12\relax}%
\providecommand \@@startlink[1]{}%
\providecommand \@@endlink[0]{}%
\providecommand \url  [0]{\begingroup\@sanitize@url \@url }%
\providecommand \@url [1]{\endgroup\@href {#1}{\urlprefix }}%
\providecommand \urlprefix  [0]{URL }%
\providecommand \Eprint [0]{\href }%
\providecommand \doibase [0]{http://dx.doi.org/}%
\providecommand \selectlanguage [0]{\@gobble}%
\providecommand \bibinfo  [0]{\@secondoftwo}%
\providecommand \bibfield  [0]{\@secondoftwo}%
\providecommand \translation [1]{[#1]}%
\providecommand \BibitemOpen [0]{}%
\providecommand \bibitemStop [0]{}%
\providecommand \bibitemNoStop [0]{.\EOS\space}%
\providecommand \EOS [0]{\spacefactor3000\relax}%
\providecommand \BibitemShut  [1]{\csname bibitem#1\endcsname}%
\let\auto@bib@innerbib\@empty
\bibitem [{\citenamefont {Valle}\ \emph {et~al.}(2014)\citenamefont {Valle},
  \citenamefont {Milotti}, \citenamefont {Ejlli}, \citenamefont {Messineo},
  \citenamefont {Piemontese}, \citenamefont {Zavattini}, \citenamefont
  {Gastaldi}, \citenamefont {Pengo},\ and\ \citenamefont
  {Ruoso}}]{Della_Valle_2014}%
  \BibitemOpen
  \bibfield  {author} {\bibinfo {author} {\bibfnamefont {F.~D.}\ \bibnamefont
  {Valle}}, \bibinfo {author} {\bibfnamefont {E.}~\bibnamefont {Milotti}},
  \bibinfo {author} {\bibfnamefont {A.}~\bibnamefont {Ejlli}}, \bibinfo
  {author} {\bibfnamefont {G.}~\bibnamefont {Messineo}}, \bibinfo {author}
  {\bibfnamefont {L.}~\bibnamefont {Piemontese}}, \bibinfo {author}
  {\bibfnamefont {G.}~\bibnamefont {Zavattini}}, \bibinfo {author}
  {\bibfnamefont {U.}~\bibnamefont {Gastaldi}}, \bibinfo {author}
  {\bibfnamefont {R.}~\bibnamefont {Pengo}}, \ and\ \bibinfo {author}
  {\bibfnamefont {G.}~\bibnamefont {Ruoso}},\ }\href {\doibase
  10.1103/physrevd.90.092003} {\bibfield  {journal} {\bibinfo  {journal}
  {Physical Review D}\ }\textbf {\bibinfo {volume} {90}} (\bibinfo {year}
  {2014}),\ 10.1103/physrevd.90.092003}\BibitemShut {NoStop}%
\bibitem [{\citenamefont {Ejlli}\ \emph {et~al.}(2020)\citenamefont {Ejlli},
  \citenamefont {{Della Valle}}, \citenamefont {Gastaldi}, \citenamefont
  {Messineo}, \citenamefont {Pengo}, \citenamefont {Ruoso},\ and\ \citenamefont
  {Zavattini}}]{EJLLI20201}%
  \BibitemOpen
  \bibfield  {author} {\bibinfo {author} {\bibfnamefont {A.}~\bibnamefont
  {Ejlli}}, \bibinfo {author} {\bibfnamefont {F.}~\bibnamefont {{Della
  Valle}}}, \bibinfo {author} {\bibfnamefont {U.}~\bibnamefont {Gastaldi}},
  \bibinfo {author} {\bibfnamefont {G.}~\bibnamefont {Messineo}}, \bibinfo
  {author} {\bibfnamefont {R.}~\bibnamefont {Pengo}}, \bibinfo {author}
  {\bibfnamefont {G.}~\bibnamefont {Ruoso}}, \ and\ \bibinfo {author}
  {\bibfnamefont {G.}~\bibnamefont {Zavattini}},\ }\href {\doibase
  https://doi.org/10.1016/j.physrep.2020.06.001} {\bibfield  {journal}
  {\bibinfo  {journal} {Physics Reports}\ }\textbf {\bibinfo {volume} {871}},\
  \bibinfo {pages} {1} (\bibinfo {year} {2020})}\BibitemShut {NoStop}%
\bibitem [{\citenamefont {Ataman}(2023)}]{Ataman:2023gin}%
  \BibitemOpen
  \bibfield  {author} {\bibinfo {author} {\bibfnamefont {S.}~\bibnamefont
  {Ataman}},\ }\href {\doibase 10.1088/1742-6596/2494/1/012019} {\bibfield
  {journal} {\bibinfo  {journal} {J. Phys. Conf. Ser.}\ }\textbf {\bibinfo
  {volume} {2494}},\ \bibinfo {pages} {012019} (\bibinfo {year}
  {2023})}\BibitemShut {NoStop}%
\bibitem [{\citenamefont {{Ataman}}\ and\ \citenamefont
  {{Nakamiya}}(2025)}]{Ataman:2025}%
  \BibitemOpen
  \bibfield  {author} {\bibinfo {author} {\bibfnamefont {S.}~\bibnamefont
  {{Ataman}}}\ and\ \bibinfo {author} {\bibfnamefont {Y.}~\bibnamefont
  {{Nakamiya}}},\ }\href {\doibase 10.1088/1402-4896/ade5d7} {\bibfield
  {journal} {\bibinfo  {journal} {J Physica Scripta}\ }\textbf {\bibinfo
  {volume} {100}},\ \bibinfo {eid} {075537} (\bibinfo {year}
  {2025})}\BibitemShut {NoStop}%
\bibitem [{\citenamefont {Shen}\ \emph {et~al.}(2018)\citenamefont {Shen},
  \citenamefont {Bu}, \citenamefont {Xu}, \citenamefont {Xu}, \citenamefont
  {Ji}, \citenamefont {Li},\ and\ \citenamefont {Xu}}]{Shen_2018}%
  \BibitemOpen
  \bibfield  {author} {\bibinfo {author} {\bibfnamefont {B.}~\bibnamefont
  {Shen}}, \bibinfo {author} {\bibfnamefont {Z.}~\bibnamefont {Bu}}, \bibinfo
  {author} {\bibfnamefont {J.}~\bibnamefont {Xu}}, \bibinfo {author}
  {\bibfnamefont {T.}~\bibnamefont {Xu}}, \bibinfo {author} {\bibfnamefont
  {L.}~\bibnamefont {Ji}}, \bibinfo {author} {\bibfnamefont {R.}~\bibnamefont
  {Li}}, \ and\ \bibinfo {author} {\bibfnamefont {Z.}~\bibnamefont {Xu}},\
  }\href {\doibase 10.1088/1361-6587/aaa7fb} {\bibfield  {journal} {\bibinfo
  {journal} {Plasma Physics and Controlled Fusion}\ }\textbf {\bibinfo {volume}
  {60}},\ \bibinfo {pages} {044002} (\bibinfo {year} {2018})}\BibitemShut
  {NoStop}%
\bibitem [{\citenamefont {Heisenberg}\ and\ \citenamefont
  {Euler}(1936)}]{heisenberg2006consequences}%
  \BibitemOpen
  \bibfield  {author} {\bibinfo {author} {\bibfnamefont {W.}~\bibnamefont
  {Heisenberg}}\ and\ \bibinfo {author} {\bibfnamefont {H.}~\bibnamefont
  {Euler}},\ }\href {\doibase 10.1007/BF01343663} {\enquote {\bibinfo {title}
  {Folgerungen aus der diracschen theorie des positrons},}\ } (\bibinfo {year}
  {1936}),\ \bibinfo {note} {translated in English: {Consequences of Diracs
  Theory of the Positron}, by W. Korolevski and H. Kleinert,
  arXiv:physics/0605038}\BibitemShut {NoStop}%
\bibitem [{\citenamefont {Born}\ \emph {et~al.}(1934)\citenamefont {Born},
  \citenamefont {Infeld},\ and\ \citenamefont
  {Fowler}}]{doi:10.1098/rspa.1934.0059}%
  \BibitemOpen
  \bibfield  {author} {\bibinfo {author} {\bibfnamefont {M.}~\bibnamefont
  {Born}}, \bibinfo {author} {\bibfnamefont {L.}~\bibnamefont {Infeld}}, \ and\
  \bibinfo {author} {\bibfnamefont {R.~H.}\ \bibnamefont {Fowler}},\ }\href
  {\doibase 10.1098/rspa.1934.0059} {\bibfield  {journal} {\bibinfo  {journal}
  {Proceedings of the Royal Society of London. Series A, Containing Papers of a
  Mathematical and Physical Character}\ }\textbf {\bibinfo {volume} {144}},\
  \bibinfo {pages} {425} (\bibinfo {year} {1934})},\ \Eprint
  {http://arxiv.org/abs/https://royalsocietypublishing.org/doi/pdf/10.1098/rspa.1934.0059}
  {https://royalsocietypublishing.org/doi/pdf/10.1098/rspa.1934.0059}
  \BibitemShut {NoStop}%
\bibitem [{\citenamefont {Fradkin}\ and\ \citenamefont
  {Tseytlin}(1985)}]{FRADKIN1985123}%
  \BibitemOpen
  \bibfield  {author} {\bibinfo {author} {\bibfnamefont {E.}~\bibnamefont
  {Fradkin}}\ and\ \bibinfo {author} {\bibfnamefont {A.}~\bibnamefont
  {Tseytlin}},\ }\href {\doibase https://doi.org/10.1016/0370-2693(85)90205-9}
  {\bibfield  {journal} {\bibinfo  {journal} {Physics Letters B}\ }\textbf
  {\bibinfo {volume} {163}},\ \bibinfo {pages} {123} (\bibinfo {year}
  {1985})}\BibitemShut {NoStop}%
\bibitem [{\citenamefont {Gibbons}(2001)}]{Gibbons_2001}%
  \BibitemOpen
  \bibfield  {author} {\bibinfo {author} {\bibfnamefont {G.~W.}\ \bibnamefont
  {Gibbons}},\ }in\ \href {\doibase 10.1063/1.1419338} {\emph {\bibinfo
  {booktitle} {{AIP} Conference Proceedings}}}\ (\bibinfo  {publisher}
  {{AIP}},\ \bibinfo {year} {2001})\BibitemShut {NoStop}%
\bibitem [{\citenamefont {Jackson}(1999)}]{jackson_classical_1999}%
  \BibitemOpen
  \bibfield  {author} {\bibinfo {author} {\bibfnamefont {J.~D.}\ \bibnamefont
  {Jackson}},\ }\href@noop {} {\emph {\bibinfo {title} {Classical
  electrodynamics}}},\ \bibinfo {edition} {3rd}\ ed.\ (\bibinfo  {publisher}
  {Wiley},\ \bibinfo {address} {New York, {NY}},\ \bibinfo {year}
  {1999})\BibitemShut {NoStop}%
\bibitem [{\citenamefont {Bialynicki-Birula}(1983)}]{Bialynicki-Birula1983}%
  \BibitemOpen
  \bibfield  {author} {\bibinfo {author} {\bibfnamefont {I.}~\bibnamefont
  {Bialynicki-Birula}},\ }\enquote {\bibinfo {title} {Nonlinear
  electrodynamics: Variations on a theme by born and infeld},}\ in\ \href@noop
  {} {\emph {\bibinfo {booktitle} {Quantum theory of particles and fields :
  birthday Volume dedicated to Jan Lopuszanski}}},\ \bibinfo {editor} {edited
  by\ \bibinfo {editor} {\bibfnamefont {B.}~\bibnamefont {Jancewicz}}\ and\
  \bibinfo {editor} {\bibfnamefont {J.}~\bibnamefont {Lukierski}}}\ (\bibinfo
  {publisher} {{WORLD} {SCIENTIFIC}},\ \bibinfo {year} {1983})\ pp.\ \bibinfo
  {pages} {31--48}\BibitemShut {NoStop}%
\bibitem [{\citenamefont
  {Bialynicki-Birula}(1992)}]{Bialynicki-Birula:1992rcm}%
  \BibitemOpen
  \bibfield  {author} {\bibinfo {author} {\bibfnamefont {I.}~\bibnamefont
  {Bialynicki-Birula}},\ }\href@noop {} {\bibfield  {journal} {\bibinfo
  {journal} {Acta Phys. Polon. B}\ }\textbf {\bibinfo {volume} {23}},\ \bibinfo
  {pages} {553} (\bibinfo {year} {1992})}\BibitemShut {NoStop}%
\bibitem [{\citenamefont {Bandos}\ \emph {et~al.}(2020)\citenamefont {Bandos},
  \citenamefont {Lechner}, \citenamefont {Sorokin},\ and\ \citenamefont
  {Townsend}}]{Bandos_2020}%
  \BibitemOpen
  \bibfield  {author} {\bibinfo {author} {\bibfnamefont {I.}~\bibnamefont
  {Bandos}}, \bibinfo {author} {\bibfnamefont {K.}~\bibnamefont {Lechner}},
  \bibinfo {author} {\bibfnamefont {D.}~\bibnamefont {Sorokin}}, \ and\
  \bibinfo {author} {\bibfnamefont {P.~K.}\ \bibnamefont {Townsend}},\ }\href
  {\doibase 10.1103/physrevd.102.121703} {\bibfield  {journal} {\bibinfo
  {journal} {Physical Review D}\ }\textbf {\bibinfo {volume} {102}} (\bibinfo
  {year} {2020}),\ 10.1103/physrevd.102.121703}\BibitemShut {NoStop}%
\bibitem [{\citenamefont {Shi}\ \emph {et~al.}(2023)\citenamefont {Shi},
  \citenamefont {Tan},\ and\ \citenamefont {Wang}}]{shi2023transverse}%
  \BibitemOpen
  \bibfield  {author} {\bibinfo {author} {\bibfnamefont {Y.}~\bibnamefont
  {Shi}}, \bibinfo {author} {\bibfnamefont {Q.}~\bibnamefont {Tan}}, \ and\
  \bibinfo {author} {\bibfnamefont {T.}~\bibnamefont {Wang}},\ }\href@noop {}
  {\enquote {\bibinfo {title} {Transverse electric waves in
  bandos-lechner-sorokin-townsend nonlinear electrodynamics},}\ } (\bibinfo
  {year} {2023}),\ \Eprint {http://arxiv.org/abs/2312.16031} {arXiv:2312.16031
  [physics.class-ph]} \BibitemShut {NoStop}%
\bibitem [{\citenamefont {Bandos}\ \emph {et~al.}(2021)\citenamefont {Bandos},
  \citenamefont {Lechner}, \citenamefont {Sorokin},\ and\ \citenamefont
  {Townsend}}]{Bandos_2021}%
  \BibitemOpen
  \bibfield  {author} {\bibinfo {author} {\bibfnamefont {I.}~\bibnamefont
  {Bandos}}, \bibinfo {author} {\bibfnamefont {K.}~\bibnamefont {Lechner}},
  \bibinfo {author} {\bibfnamefont {D.}~\bibnamefont {Sorokin}}, \ and\
  \bibinfo {author} {\bibfnamefont {P.~K.}\ \bibnamefont {Townsend}},\ }\href
  {\doibase 10.1007/jhep03(2021)022} {\bibfield  {journal} {\bibinfo  {journal}
  {Journal of High Energy Physics}\ }\textbf {\bibinfo {volume} {2021}}
  (\bibinfo {year} {2021}),\ 10.1007/jhep03(2021)022}\BibitemShut {NoStop}%
\bibitem [{\citenamefont {Sorokin}(2022)}]{Sorokin_2022}%
  \BibitemOpen
  \bibfield  {author} {\bibinfo {author} {\bibfnamefont {D.~P.}\ \bibnamefont
  {Sorokin}},\ }\href {\doibase 10.1002/prop.202200092} {\bibfield  {journal}
  {\bibinfo  {journal} {Fortschritte der Physik}\ }\textbf {\bibinfo {volume}
  {70}},\ \bibinfo {pages} {2200092} (\bibinfo {year} {2022})}\BibitemShut
  {NoStop}%
\bibitem [{\citenamefont {Conti}\ \emph {et~al.}(2018)\citenamefont {Conti},
  \citenamefont {Iannella}, \citenamefont {Negro},\ and\ \citenamefont
  {Tateo}}]{Conti_2018}%
  \BibitemOpen
  \bibfield  {author} {\bibinfo {author} {\bibfnamefont {R.}~\bibnamefont
  {Conti}}, \bibinfo {author} {\bibfnamefont {L.}~\bibnamefont {Iannella}},
  \bibinfo {author} {\bibfnamefont {S.}~\bibnamefont {Negro}}, \ and\ \bibinfo
  {author} {\bibfnamefont {R.}~\bibnamefont {Tateo}},\ }\href {\doibase
  10.1007/jhep11(2018)007} {\bibfield  {journal} {\bibinfo  {journal} {Journal
  of High Energy Physics}\ }\textbf {\bibinfo {volume} {2018}} (\bibinfo {year}
  {2018}),\ 10.1007/jhep11(2018)007}\BibitemShut {NoStop}%
\bibitem [{\citenamefont {Ferko}\ \emph {et~al.}(2022)\citenamefont {Ferko},
  \citenamefont {Smith},\ and\ \citenamefont {Tartaglino}}]{Ferko_2022}%
  \BibitemOpen
  \bibfield  {author} {\bibinfo {author} {\bibfnamefont {C.}~\bibnamefont
  {Ferko}}, \bibinfo {author} {\bibfnamefont {L.}~\bibnamefont {Smith}}, \ and\
  \bibinfo {author} {\bibfnamefont {G.}~\bibnamefont {Tartaglino}},\ }\href
  {\doibase 10.21468/scipostphys.13.2.012} {\bibfield  {journal} {\bibinfo
  {journal} {{SciPost} Physics}\ }\textbf {\bibinfo {volume} {13}} (\bibinfo
  {year} {2022}),\ 10.21468/scipostphys.13.2.012}\BibitemShut {NoStop}%
\bibitem [{\citenamefont {Babaei-Aghbolagh}\ \emph {et~al.}(2022)\citenamefont
  {Babaei-Aghbolagh}, \citenamefont {{Babaei Velni}}, \citenamefont {Yekta},\
  and\ \citenamefont {Mohammadzadeh}}]{BABAEIAGHBOLAGH2022137079}%
  \BibitemOpen
  \bibfield  {author} {\bibinfo {author} {\bibfnamefont {H.}~\bibnamefont
  {Babaei-Aghbolagh}}, \bibinfo {author} {\bibfnamefont {K.}~\bibnamefont
  {{Babaei Velni}}}, \bibinfo {author} {\bibfnamefont {D.~M.}\ \bibnamefont
  {Yekta}}, \ and\ \bibinfo {author} {\bibfnamefont {H.}~\bibnamefont
  {Mohammadzadeh}},\ }\href {\doibase
  https://doi.org/10.1016/j.physletb.2022.137079} {\bibfield  {journal}
  {\bibinfo  {journal} {Physics Letters B}\ }\textbf {\bibinfo {volume}
  {829}},\ \bibinfo {pages} {137079} (\bibinfo {year} {2022})}\BibitemShut
  {NoStop}%
\bibitem [{\citenamefont {Ferko}\ \emph {et~al.}(2023)\citenamefont {Ferko},
  \citenamefont {Kuzenko}, \citenamefont {Smith},\ and\ \citenamefont
  {Tartaglino-Mazzucchelli}}]{ferko2023dualityinvariant}%
  \BibitemOpen
  \bibfield  {author} {\bibinfo {author} {\bibfnamefont {C.}~\bibnamefont
  {Ferko}}, \bibinfo {author} {\bibfnamefont {S.~M.}\ \bibnamefont {Kuzenko}},
  \bibinfo {author} {\bibfnamefont {L.}~\bibnamefont {Smith}}, \ and\ \bibinfo
  {author} {\bibfnamefont {G.}~\bibnamefont {Tartaglino-Mazzucchelli}},\ }\href
  {\doibase 10.1103/PhysRevD.108.106021} {\bibfield  {journal} {\bibinfo
  {journal} {Phys. Rev. D}\ }\textbf {\bibinfo {volume} {108}},\ \bibinfo
  {pages} {106021} (\bibinfo {year} {2023})}\BibitemShut {NoStop}%
\bibitem [{\citenamefont {Ferraro}(2007)}]{Ferraro_2007}%
  \BibitemOpen
  \bibfield  {author} {\bibinfo {author} {\bibfnamefont {R.}~\bibnamefont
  {Ferraro}},\ }\href {\doibase 10.1103/physrevlett.99.230401} {\bibfield
  {journal} {\bibinfo  {journal} {Physical Review Letters}\ }\textbf {\bibinfo
  {volume} {99}} (\bibinfo {year} {2007}),\
  10.1103/physrevlett.99.230401}\BibitemShut {NoStop}%
\bibitem [{\citenamefont {Manojlovic}\ \emph {et~al.}(2020)\citenamefont
  {Manojlovic}, \citenamefont {Perlick},\ and\ \citenamefont
  {Potting}}]{MANOJLOVIC2020168303}%
  \BibitemOpen
  \bibfield  {author} {\bibinfo {author} {\bibfnamefont {N.}~\bibnamefont
  {Manojlovic}}, \bibinfo {author} {\bibfnamefont {V.}~\bibnamefont {Perlick}},
  \ and\ \bibinfo {author} {\bibfnamefont {R.}~\bibnamefont {Potting}},\ }\href
  {\doibase https://doi.org/10.1016/j.aop.2020.168303} {\bibfield  {journal}
  {\bibinfo  {journal} {Annals of Physics}\ }\textbf {\bibinfo {volume}
  {422}},\ \bibinfo {pages} {168303} (\bibinfo {year} {2020})}\BibitemShut
  {NoStop}%
\bibitem [{\citenamefont {Russo}\ and\ \citenamefont
  {Townsend}(2023)}]{Russo_2023}%
  \BibitemOpen
  \bibfield  {author} {\bibinfo {author} {\bibfnamefont {J.~G.}\ \bibnamefont
  {Russo}}\ and\ \bibinfo {author} {\bibfnamefont {P.~K.}\ \bibnamefont
  {Townsend}},\ }\href {\doibase 10.1007/jhep01(2023)039} {\bibfield  {journal}
  {\bibinfo  {journal} {Journal of High Energy Physics}\ }\textbf {\bibinfo
  {volume} {2023}} (\bibinfo {year} {2023}),\
  10.1007/jhep01(2023)039}\BibitemShut {NoStop}%
\bibitem [{\citenamefont {Neves}\ \emph {et~al.}(2023)\citenamefont {Neves},
  \citenamefont {Gaete}, \citenamefont {Ospedal},\ and\ \citenamefont
  {Helayël-Neto}}]{Neves_2023}%
  \BibitemOpen
  \bibfield  {author} {\bibinfo {author} {\bibfnamefont {M.~J.}\ \bibnamefont
  {Neves}}, \bibinfo {author} {\bibfnamefont {P.}~\bibnamefont {Gaete}},
  \bibinfo {author} {\bibfnamefont {L.~P.~R.}\ \bibnamefont {Ospedal}}, \ and\
  \bibinfo {author} {\bibfnamefont {J.~A.}\ \bibnamefont {Helayël-Neto}},\
  }\href {\doibase 10.1103/physrevd.107.075019} {\bibfield  {journal} {\bibinfo
   {journal} {Physical Review D}\ }\textbf {\bibinfo {volume} {107}} (\bibinfo
  {year} {2023}),\ 10.1103/physrevd.107.075019}\BibitemShut {NoStop}%
\bibitem [{\citenamefont {Mezincescu}\ \emph {et~al.}(2024)\citenamefont
  {Mezincescu}, \citenamefont {Russo},\ and\ \citenamefont
  {Townsend}}]{mezincescu2024hamiltonian}%
  \BibitemOpen
  \bibfield  {author} {\bibinfo {author} {\bibfnamefont {L.}~\bibnamefont
  {Mezincescu}}, \bibinfo {author} {\bibfnamefont {J.~G.}\ \bibnamefont
  {Russo}}, \ and\ \bibinfo {author} {\bibfnamefont {P.~K.}\ \bibnamefont
  {Townsend}},\ }\href {\doibase 10.1007/JHEP02(2024)186} {\bibfield  {journal}
  {\bibinfo  {journal} {Journal of High Energy Physics}\ }\textbf {\bibinfo
  {volume} {02}},\ \bibinfo {pages} {186} (\bibinfo {year} {2024})}\BibitemShut
  {NoStop}%
\bibitem [{\citenamefont {Neves}\ \emph {et~al.}(2021)\citenamefont {Neves},
  \citenamefont {de~Oliveira}, \citenamefont {Ospedal},\ and\ \citenamefont
  {Helay\"el-Neto}}]{PhysRevD.104.015006}%
  \BibitemOpen
  \bibfield  {author} {\bibinfo {author} {\bibfnamefont {M.~J.}\ \bibnamefont
  {Neves}}, \bibinfo {author} {\bibfnamefont {J.~B.}\ \bibnamefont
  {de~Oliveira}}, \bibinfo {author} {\bibfnamefont {L.~P.~R.}\ \bibnamefont
  {Ospedal}}, \ and\ \bibinfo {author} {\bibfnamefont {J.~A.}\ \bibnamefont
  {Helay\"el-Neto}},\ }\href {\doibase 10.1103/PhysRevD.104.015006} {\bibfield
  {journal} {\bibinfo  {journal} {Phys. Rev. D}\ }\textbf {\bibinfo {volume}
  {104}},\ \bibinfo {pages} {015006} (\bibinfo {year} {2021})}\BibitemShut
  {NoStop}%
\bibitem [{\citenamefont {Lechner}\ \emph {et~al.}(2022)\citenamefont
  {Lechner}, \citenamefont {Marchetti}, \citenamefont {Sainaghi},\ and\
  \citenamefont {Sorokin}}]{Lechner_2022}%
  \BibitemOpen
  \bibfield  {author} {\bibinfo {author} {\bibfnamefont {K.}~\bibnamefont
  {Lechner}}, \bibinfo {author} {\bibfnamefont {P.}~\bibnamefont {Marchetti}},
  \bibinfo {author} {\bibfnamefont {A.}~\bibnamefont {Sainaghi}}, \ and\
  \bibinfo {author} {\bibfnamefont {D.}~\bibnamefont {Sorokin}},\ }\href
  {\doibase 10.1103/physrevd.106.016009} {\bibfield  {journal} {\bibinfo
  {journal} {Physical Review D}\ }\textbf {\bibinfo {volume} {106}} (\bibinfo
  {year} {2022}),\ 10.1103/physrevd.106.016009}\BibitemShut {NoStop}%
\bibitem [{\citenamefont {Shi}\ and\ \citenamefont
  {Wang}(2025)}]{SHI_Blackbody}%
  \BibitemOpen
  \bibfield  {author} {\bibinfo {author} {\bibfnamefont {Y.}~\bibnamefont
  {Shi}}\ and\ \bibinfo {author} {\bibfnamefont {T.}~\bibnamefont {Wang}},\
  }\href {\doibase 10.3969/j.issn.1000-5641.2025.03.002} {\bibfield  {journal}
  {\bibinfo  {journal} {Journal of East China Normal University (Natural
  Science)}\ }\textbf {\bibinfo {volume} {2025}},\ \bibinfo {eid} {13}
  (\bibinfo {year} {2025})}\BibitemShut {NoStop}%
\bibitem [{\citenamefont {Pérez-García}\ \emph {et~al.}(2023)\citenamefont
  {Pérez-García}, \citenamefont {Pérez~Martínez},\ and\ \citenamefont
  {Rodríguez~Querts}}]{P_rez_Garc_a_2023}%
  \BibitemOpen
  \bibfield  {author} {\bibinfo {author} {\bibfnamefont {M.~A.}\ \bibnamefont
  {Pérez-García}}, \bibinfo {author} {\bibfnamefont {A.}~\bibnamefont
  {Pérez~Martínez}}, \ and\ \bibinfo {author} {\bibfnamefont
  {E.}~\bibnamefont {Rodríguez~Querts}},\ }\href {\doibase
  10.1140/epjc/s10052-023-11902-3} {\bibfield  {journal} {\bibinfo  {journal}
  {The European Physical Journal C}\ }\textbf {\bibinfo {volume} {83}}
  (\bibinfo {year} {2023}),\ 10.1140/epjc/s10052-023-11902-3}\BibitemShut
  {NoStop}%
\bibitem [{\citenamefont {Fedotov}\ \emph {et~al.}(2023)\citenamefont
  {Fedotov}, \citenamefont {Ilderton}, \citenamefont {Karbstein}, \citenamefont
  {King}, \citenamefont {Seipt}, \citenamefont {Taya},\ and\ \citenamefont
  {Torgrimsson}}]{Fedotov_2023}%
  \BibitemOpen
  \bibfield  {author} {\bibinfo {author} {\bibfnamefont {A.}~\bibnamefont
  {Fedotov}}, \bibinfo {author} {\bibfnamefont {A.}~\bibnamefont {Ilderton}},
  \bibinfo {author} {\bibfnamefont {F.}~\bibnamefont {Karbstein}}, \bibinfo
  {author} {\bibfnamefont {B.}~\bibnamefont {King}}, \bibinfo {author}
  {\bibfnamefont {D.}~\bibnamefont {Seipt}}, \bibinfo {author} {\bibfnamefont
  {H.}~\bibnamefont {Taya}}, \ and\ \bibinfo {author} {\bibfnamefont
  {G.}~\bibnamefont {Torgrimsson}},\ }\href {\doibase
  10.1016/j.physrep.2023.01.003} {\bibfield  {journal} {\bibinfo  {journal}
  {Physics Reports}\ }\textbf {\bibinfo {volume} {1010}},\ \bibinfo {pages} {1}
  (\bibinfo {year} {2023})}\BibitemShut {NoStop}%
\end{thebibliography}%

\end{document}